\pdfoutput=1

\documentclass[11pt]{article}

\usepackage[preprint]{acl}

\usepackage{times}
\usepackage{latexsym}
\usepackage{listings}

\usepackage{multirow}
\usepackage[T1]{fontenc}

\usepackage[utf8]{inputenc}
\usepackage[normalem]{ulem}
\usepackage{microtype}

\usepackage{inconsolata}
\usepackage{longtable}

\usepackage{graphicx}
\usepackage{tabularx}
\usepackage{amsmath}
\usepackage{amssymb}
\usepackage{algorithm}
\usepackage{algpseudocode}
\usepackage{subcaption}
\usepackage{booktabs}
\usepackage{float}
\usepackage{multirow}
\usepackage{graphicx}
\usepackage{makecell}
\usepackage{amssymb}
\usepackage{pifont}
\usepackage{xcolor}
\usepackage{colortbl}
\usepackage{amsmath}
\newcolumntype{M}{>{\centering\arraybackslash}m{2cm}} 
\newcolumntype{C}{>{\centering\arraybackslash}X} 
\usepackage{tabularray}
\usepackage{url}

\UseTblrLibrary{diagbox}
\usepackage{fontawesome}
\definecolor{mygray}{gray}{.9}
\makeatletter
\renewcommand\@makefntext[1]{%
    \setlength{\hangindent}{1em} 
    \noindent
    \hb@xt@1em{\hss\@makefnmark}#1} 
\makeatother

\title{BlenderLLM: Training Large Language Models for Computer-Aided Design with Self-improvement}

\author{
\textbf{Yuhao Du\textsuperscript{$\alpha$}}, 
\textbf{Shunian Chen\textsuperscript{$\alpha$}}, 
\textbf{Wenbo Zan\textsuperscript{$\alpha$}}, 
\textbf{Peizhao Li\textsuperscript{$\alpha$}}, 
\textbf{Mingxuan Wang\textsuperscript{$\alpha$}}, 
\\
\textbf{Dingjie Song\textsuperscript{$\alpha$}},
\textbf{Bo Li\textsuperscript{$\gamma$}}, 
\textbf{Yan Hu\textsuperscript{$\alpha$}}, 
\textbf{Benyou Wang\textsuperscript{$\alpha$}}* \\
\textsuperscript{$\alpha$}The Chinese University of Hong Kong, Shenzhen \\
\textsuperscript{$\gamma$}Information Technology Research Institute, China Tower \\
\small{\textbf{\textsuperscript{*}Correspondence:} \href{mailto:wangbenyou@cuhk.edu.cn}{wangbenyou@cuhk.edu.cn}}
}

\usepackage{color}
\definecolor{Alto}{rgb}{0.862,0.862,0.862}

\usepackage{inconsolata}

\begin{document}
\maketitle

\begin{abstract}
The application of Large Language Models (LLMs) in Computer-Aided Design (CAD) remains an underexplored area, despite their remarkable advancements in other domains. In this paper, we present \textbf{BlenderLLM}, a novel framework for training LLMs specifically for CAD tasks leveraging a self-improvement methodology. To support this, we developed a bespoke training dataset, \textbf{BlendNet}, and introduced a comprehensive evaluation suite, \textbf{CADBench}. Our results reveal that existing models demonstrate significant limitations in generating accurate CAD scripts. However, through minimal instruction-based fine-tuning and iterative self-improvement, BlenderLLM significantly surpasses these models in both functionality and accuracy of CAD script generation. This research establishes a strong foundation for the application of LLMs in CAD while demonstrating the transformative potential of self-improving models in advancing CAD automation. We encourage further exploration and adoption of these methodologies to drive innovation in the field. The dataset, model, benchmark, and source code are publicly available at \url{https://github.com/FreedomIntelligence/BlenderLLM}
\end{abstract}

\section{Introduction}

CAD is extensively used in industries such as automotive, aerospace, manufacturing, and architecture for 3D design \cite{heesom2004trends, pottmann2005industrial, susic2017application}. Despite its widespread application, the effective use of CAD often demands specialized skills and substantial training, making the design process both labor-intensive and time-consuming. Tasks like parameter adjustments and model validation require considerable human effort, leading to increased project costs and slowing down rapid iteration and innovation \citep{kreis2020cad}.

Large language models (LLMs) have experienced rapid advancements in recent years, particularly in architecture and training methodologies. Sophisticated models such as \texttt{GPT-4}~\citep{openai2023gpt4} have demonstrated human-like performance on a variety of tasks. Their ability to generate coherent and contextually relevant text has made them valuable across numerous applications, including potentially transforming the way CAD tasks are approached.

\setlength{\belowcaptionskip}{-13pt}

\begin{figure}[t]
    \centering
    \includegraphics[width=0.5\textwidth]{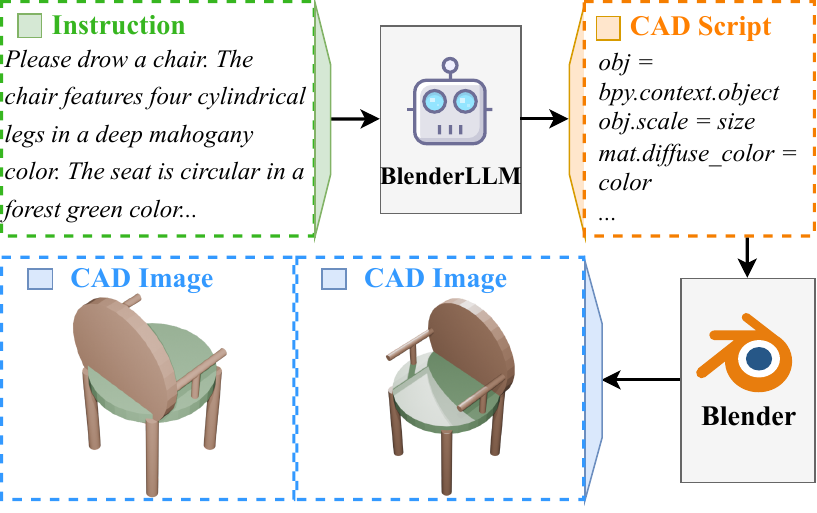}
    \caption{Illustrative Instances}
    \label{fig:Problem Definition}
\end{figure}

\paragraph{Problem Definition}
This paper addresses the challenge of reducing the manual workload associated with CAD design by leveraging the capabilities of LLMs. As illustrated in Figure~\ref{fig:Problem Definition}, we utilize LLMs to automate the generation of CAD scripts from natural language inputs. These scripts can be executed in Blender to create precise 3D models. By converting user instructions into executable CAD scripts, our approach streamlines the CAD process, thereby alleviating the manual workload for engineers and designers. 

\begin{table*}     
    \centering     
    \begin{tabular}{p{1cm}p{1.5cm}p{1.5cm}p{1.5cm}p{1.5cm}p{1.5cm}p{1.5cm}p{1.5cm}p{0cm}}  
        & \centering \small \textbf{BlenderLLM} & \centering \small o1-Preview & \centering \small \texttt{GPT-4o} & \centering \small GPT-4-Turbo & \centering \small Claude-3.5-Sonnet & \centering \small Gemini-1.5-Pro & \centering \small BlenderGPT &
        \\
        \centering \small \textit{Burger} & \centering \raisebox{-0.5\height}{\includegraphics[scale=0.2]{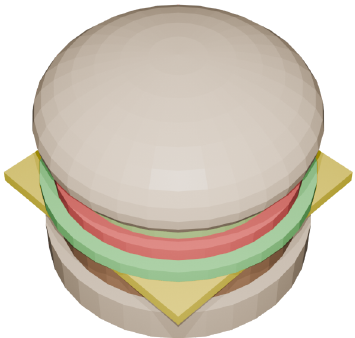}} & \centering \raisebox{-0.5\height}{\includegraphics[scale=0.2]{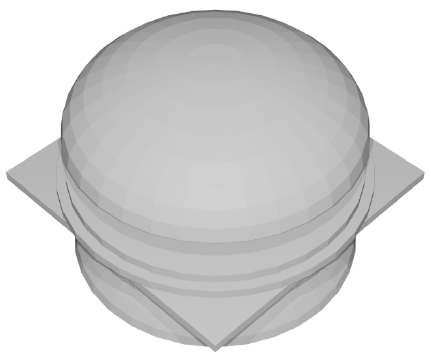}}  & \centering \raisebox{-0.5\height}{\includegraphics[scale=0.2]{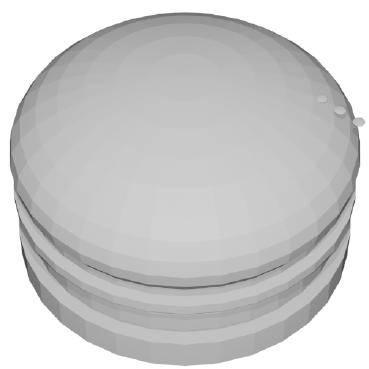}} & \centering \raisebox{-0.5\height}{\includegraphics[scale=0.17]{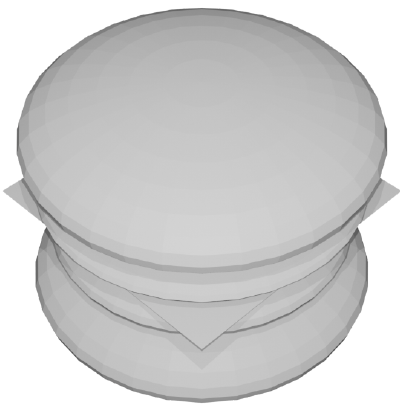}} & \centering \raisebox{-0.5\height}{\ding{55}}  & 
        \centering \raisebox{-0.5\height}{\includegraphics[scale=0.18]{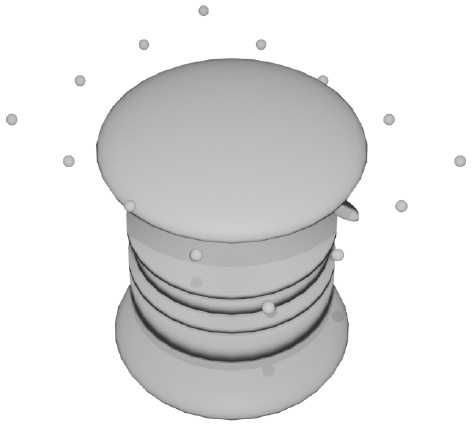}}  &
        \centering \raisebox{-0.5\height}{\includegraphics[scale=0.17]{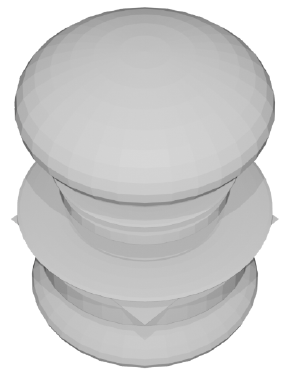}} &
        \\

        \centering \small \textit{Desk Lamp} & \centering \raisebox{-0.5\height}{\includegraphics[scale=0.2]{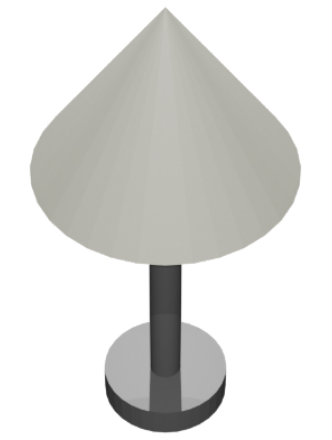}} & \centering \raisebox{-0.5\height}{\includegraphics[scale=0.17]{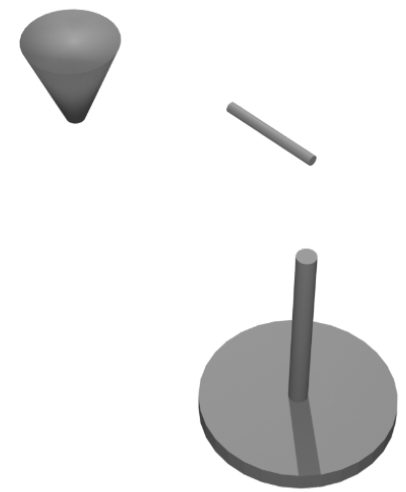}} & \centering \raisebox{-0.5\height}{\includegraphics[scale=0.2]{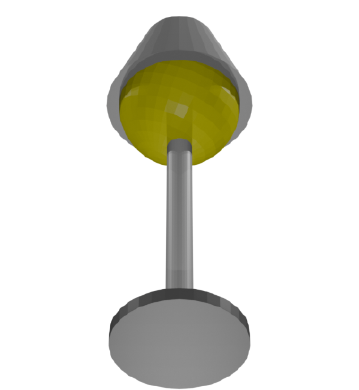}} & \centering \raisebox{-0.5\height}{\includegraphics[scale=0.2]{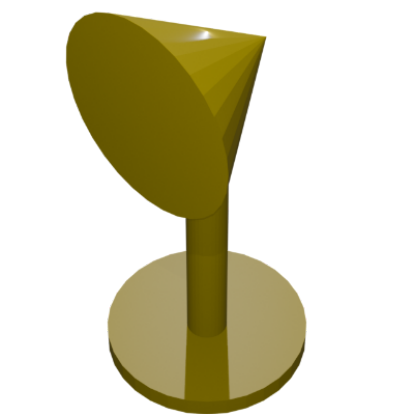}} & \centering \raisebox{-0.5\height}{\includegraphics[scale=0.2]{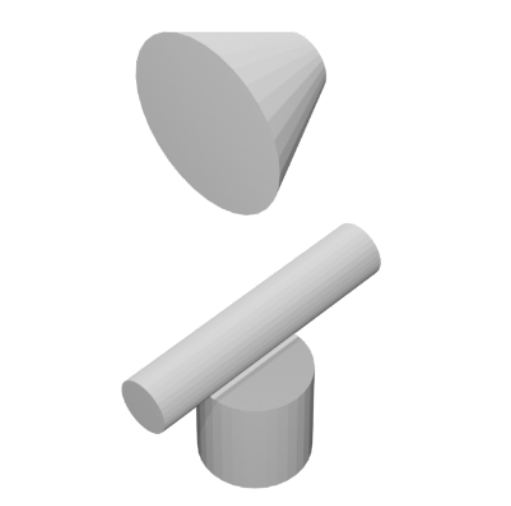}} & \centering \raisebox{-0.5\height}{\includegraphics[scale=0.18]{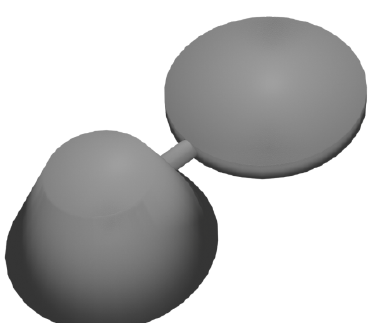}} & \centering \raisebox{-0.5\height}{\includegraphics[scale=0.2]{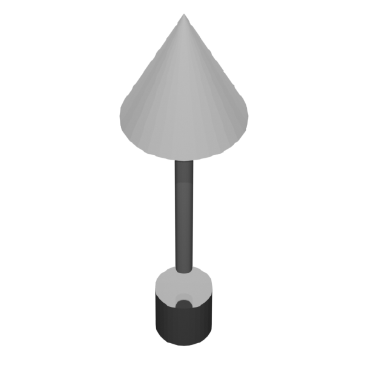}} & \\

        \centering \small \textit{Celtic Knot} & \centering \raisebox{-0.5\height}{\includegraphics[scale=0.15]{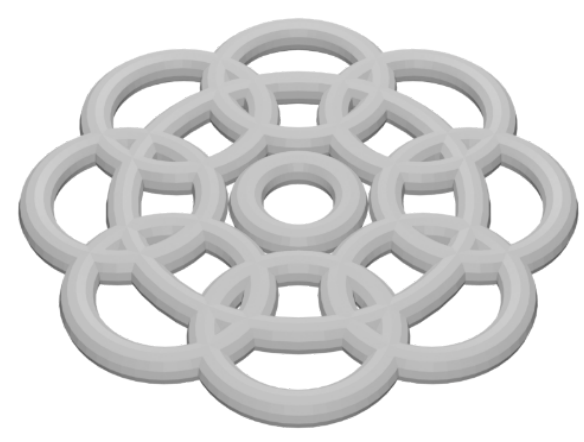}} & \centering \raisebox{-0.5\height}{\includegraphics[scale=0.15]{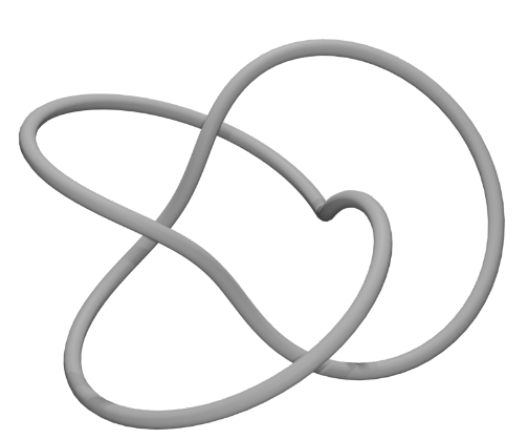}} & \centering \raisebox{-0.5\height}{\includegraphics[scale=0.15]{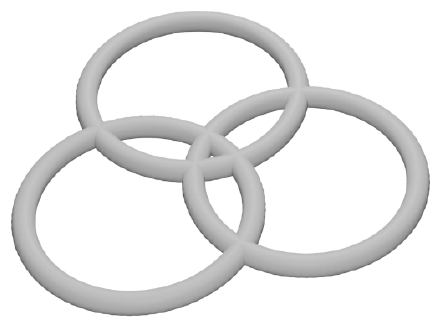}} & \centering \raisebox{-0.5\height}{\ding{55}} & \centering \raisebox{-0.5\height}{\ding{55}} & \centering \raisebox{-0.5\height}{\ding{55}} & \centering \raisebox{-0.5\height}{\ding{55}} & \\
    \end{tabular}     
    \caption{Examples of the performance of different LLMs. Note: \ding{55} means that the CAD script generated by the model result in an error during execution, thus no corresponding 3D model is generated.
}     
    \label{tab:Examples of the performance of different LLMs} 
\end{table*}

\paragraph{Challenges}
Although recent work~\citep{kreis2020cad,blendergpt2023,wu2023cadllm,CadVLM2024} has explored the application of LLM in the CAD field, several significant challenges still hinder its widespread adoption. Firstly, some work is limited by the complexity of input forms, resulting in a high threshold for use. Secondly, there is a notable shortage of high-quality, domain-specific datasets required to train models capable of capturing the intricate nuances of CAD design. Thirdly, the lack of open-source models limits accessibility, local deployment, and privacy preservation. Finally, the absence of a comprehensive evaluation framework hampers the ability to rigorously assess LLM performance in CAD applications. Addressing these challenges is critical for advancing CAD-oriented LLMs and ensuring robust, secure, and on-premises solutions.

\paragraph{Methodology}
To address the aforementioned challenges, we present a novel framework consisting of three key components that allow users to generate CAD models with natural language: \textbf{BlendNet}, a high-quality dataset comprising $8k$ samples; \textbf{BlenderLLM}, a CAD script generation model; and \textbf{CADBench}, a comprehensive benchmarking suite.
First, we construct a multi-module data generation pipeline to create BlendNet, whose samples map natural language instructions to \texttt{bpy} scripts. 
Then, we use \texttt{BlendNet} to fine-tune a model, obtaining the BlenderLLM-base. To further address the issue of data scarcity, we employ a self-improvement approach, utilizing data generated by the model itself to enhance its performance through an iterative process. 
Furthermore, we introduce a specialized benchmark, \texttt{CADBench}, an evaluation framework employing MLLM-as-judge~\citep{ge2024mllmbench} for assessing a model's capacity to generate 3D models from open-ended instructions.

Empirical evaluations demonstrate that BlenderLLM outperforms all baseline models across multiple dimensions on \texttt{CADBench}. Examples are shown in Table~\ref{tab:Examples of the performance of different LLMs}. Contributions of this paper are summarized as follows:

\begin{itemize}
    \item We introduce a high-quality dataset, \textbf{BlendNet}, comprising $8k$ diverse CAD samples, along with its data generation pipeline.
    \item We train a novel \texttt{bpy} script generation model, \textbf{BlenderLLM}, which undergoes Supervised Fine-tuning and iterative self-improvement process to achieve state-of-the-art performance.
    \item We develop a benchmarking framework, \textbf{CADBench}, to evaluate the model's ability to generate CAD scripts from user-provided instructions, enabling a systematic assessment of CAD generation capabilities.
\end{itemize}

\section{Related Work}

\subsection{Computer-Aided Design (CAD)}

CAD is a widely used technology in various industries, enabling engineers and designers to create precise digital representations of objects, offering significant advantages in precision, flexibility, and speed. Early efforts leveraged rule-based systems and simple machine learning algorithms to assist in CAD tasks \citep{chavali2008using}. Later, convolutional neural networks were used to convert 2D sketches into 3D models \citep{li2020sketch2cad}. 
However, these methods have limitations. Rule-based systems lack flexibility, while machine learning require extensive labeled data and are constrained by their training data's scope \citep{rapp2021mlcad}.

\subsection{Large Language Models for CAD}

Recent work has begun to explore how LLMs can be adapted for CAD tasks. For instance, CADGPT \citep{kapsalis2024cadgptharnessingnaturallanguage} directly parses natural language inputs into executable commands for CAD software. BlenderGPT \citep{blendergpt2023} and 3D-PREMISE \citep{Yuan20243D-PREMISE} have utilized LLMs like \texttt{GPT-4} to generate CAD scripts based on natural language prompts. Additionally, CAD-LLM \citep{wu2023cadllm} has successfully trained a T5 model for CAD sketch completion. Moreover, CadVLM \citep{CadVLM2024} introduces a multimodal approach that bridges language and vision, enabling the generation of parametric CAD sketches from both textual and visual inputs. Appendix~\ref{sec:Comparison of BlenderLLM and Recent Works} outlines the key differences between BlenderLLM and existing LLMs designed for CAD-related tasks.

\subsection{Blender}
Blender is an open-source 3D creation suite widely used in film, game development, and architectural visualization, offering a comprehensive toolset for modeling, animation, and rendering, with flexibility enhanced by its Python API (\texttt{bpy} scripts). Its advantages over other CAD software, including a lower learning curve and broader user base~\citep{blendertut, tuori2022advantages}, make it the ideal platform for CAD tasks. In our work, Blender is used for rendering CAD scripts, acting as an intermediary between the large language model outputs and the visual results.

\begin{figure*}[t]
    \centering
    \includegraphics[width=\textwidth]{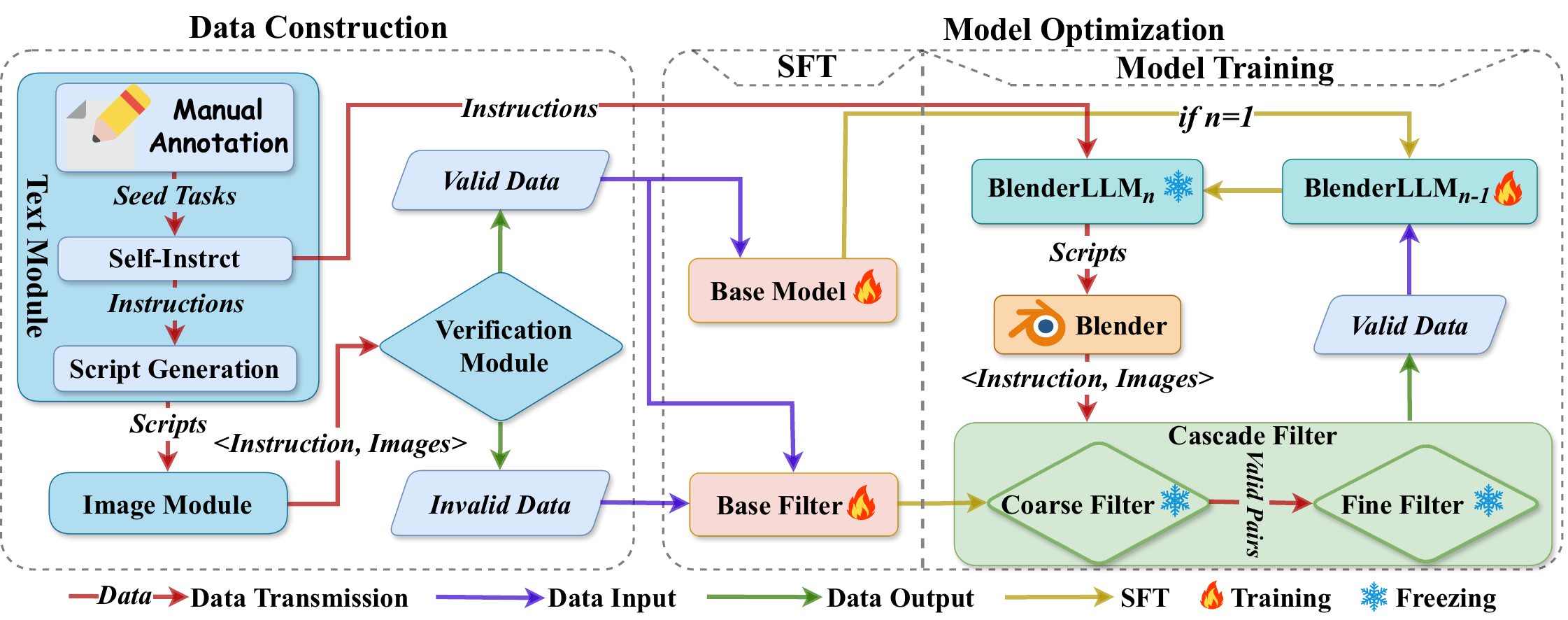}
    \caption{The Pipeline of the Methodology. In \textbf{Step I}, we utilize a multi-module pipeline to construct a high-quality training dataset and fine-tune the Base Model and Base Filter on it, establishing a foundation for the next phase. In \textbf{Step II},
    the model is fine-tuned by Self-improvement until achieving the optimal model.
}
    \label{fig:flow_diagram}
\end{figure*}

\section{Methodology}
\subsection{Data Construction}
\label{sec:data_construction}

We design and implement a multi-module pipeline for generating high-quality training data for SFT. The pipeline for data construction is illustrated in Figure~\ref{fig:flow_diagram}. The multi-module pipeline is composed of three primary components: the \textbf{Text Module}, the \textbf{Image Module}, and the \textbf{Verification Module}. The Text Module generates instructions and their corresponding \texttt{bpy} scripts. The Image Module executes these \texttt{bpy} scripts within Blender to produce images. The Verification Module ensures that the images align with the instructions, thereby validating the data quality.

\subsubsection{Text Module}
The objective of the text module is to develop diverse instructions and corresponding \texttt{bpy} scripts. 

\paragraph{Instruction Generation}
To encompass a broad range of item types, emulate various communication styles~\citep{Sims02092017}, and craft instructions with differing levels of complexity, the diversity of the instructions is categorized along three dimensions:

\begin{itemize}
\item \textbf{Object Categories:} Objects are classified into 16 categories following the Locarno classification system \citep{Locarnoclassification}, as detailed in Appendix~\ref{Category List}.
\item \textbf{Instruction Types:} We employ the Myers-Briggs Type Indicator (MBTI) \citep{myers1985guide} to create eight distinct tones for instructions, as detailed in Appendix~\ref{Instruction Type}.
\item \textbf{Complexity:} To manage the complexity of instructions, we vary their length, classifying them into five categories, as detailed in Appendix~\ref{length list}.
\end{itemize}

Based on these dimensions, we manually create a set of 135 diverse seed instructions, denoted as \( L_{\text{seed}} = \{ l_1, l_2, \ldots, l_{135} \} \), where \( l_i \) denotes the \( i^{th} \) natural language instruction. Next, we employ Self-Instruct data distillation techniques \citep{wang2022self-instruct} to expand these seed instructions into a larger dataset. In each iteration of instruction generation, we randomly sample instances from the \(L_{\text{seed}}\). These sampled instances are used to generate new instructions. Through multiple iterations, this process results in a comprehensive dataset of approximately $50k$ instructions, denoted as \( L_{\text{gen}} \). 

The distribution of both seed instructions \( L_{\text{seed}} \) and generated instructions \( L_{\text{gen}} \) by category, type, and length is illustrated in Figure~\ref{fig:diversity_data}. The detailed process is outlined in Appendix~\ref{sec:Instruction Generation Process}.

\paragraph{Script Generation} We then utilize \texttt{GPT-4o}\footnote{Model id gpt-4o-2024-08-06} to generate pairs \(\langle l_j, s_j \rangle\) based on given instructions \( l_j \). For each instruction \( l_j \in L_{\text{gen}} \), \texttt{GPT-4o} produces a corresponding script \( s_j \). The generation process ensures that each script is derived from its instruction, as detailed in Appendix~\ref{sec:Script Generation}.

\subsubsection{Image Module}

We render the scripts using Blender to generate corresponding images. For each generated 3D object, four images are captured from different angles to better capture the full view of 3D objects, resulting in \(\langle l_j, I_j \rangle\) pairs, where \( I_j = \{ i_{j,1}, i_{j,2}, i_{j,3}, i_{j,4} \} \) is the set of images.

\subsubsection{Verification Module}

We use \texttt{GPT-4o} as the validator. The model is required to determine whether the images match the instruction based on the given \(\langle l_j, I_j \rangle\) pairs, detailed instruction can be found in Appendix~\ref{sec:Validation Process}.

To verify the reliability of \texttt{GPT-4o} as the validator, we perform manual cross-validation on a portion of the data. We manually validate  \( 10k \) data points, of which 89.7\% produce consistent results with the \texttt{GPT-4o} verification, demonstrating the reliability of \texttt{GPT-4o} as a validator. Detailed cross-validation result is shown in Appendix~\ref{sec:cross_validation}.

As a result, we obtain \( 2k \) accurate \(\langle l_j, s_j \rangle\) pairs through manual verification, referred to as \texttt{BlendNet-Human}, and \( 6k \) \(\langle l_j, s_j \rangle\) pairs validated solely by \texttt{GPT-4o}, referred to as \texttt{BlendNet-GPT}. Combining these two parts, we obtain \texttt{BlendNet}.

\begin{figure*}[t]
    \centering
    \includegraphics[width=\textwidth]{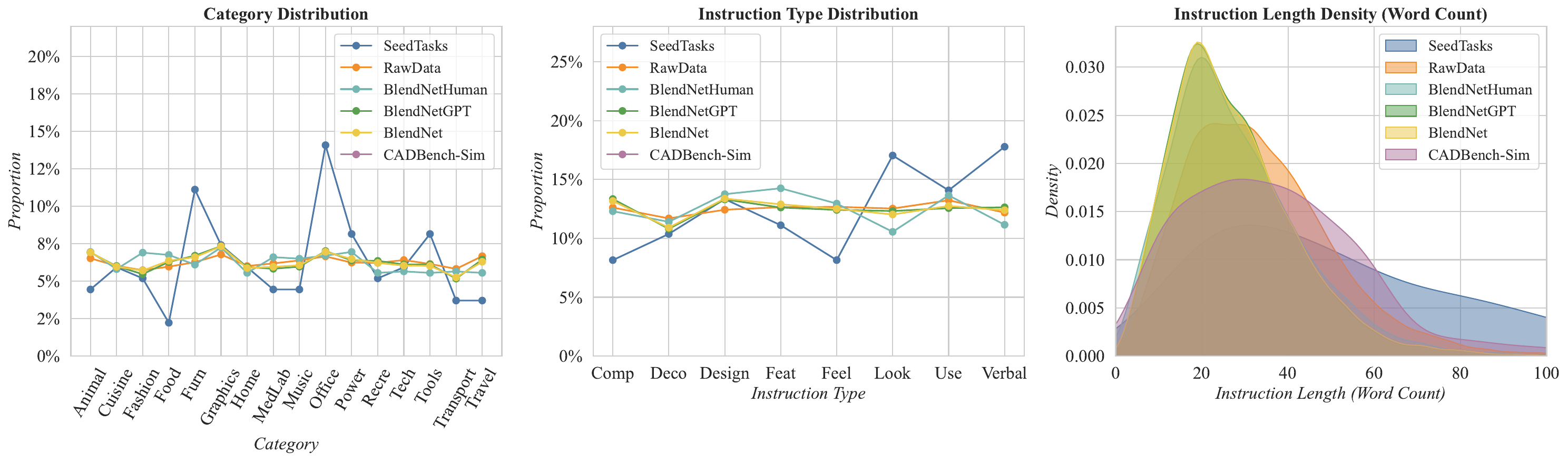}
    \caption{Diversity in Training and Evaluation Datasets. Each dataset is designed to ensure a uniform distribution across \textit{Category} and \textit{Instruction Type}, while maintaining a broad-ranging density in \textit{Instruction Length}.}
    \label{fig:diversity_data}
\end{figure*}

The diversity of \texttt{BlendNet} is illustrated in Figure~\ref{fig:diversity_data}. Additionally, we quantify the complexity of \texttt{BlendNet} tasks using three metrics: \textbf{Unit Number}, \textbf{Parameter Density}, and \textbf{Entropy} \citep{CONTERO2023101970}. More details about these metrics can be found in Appendix~\ref{sec:complexity_BlendNet}, and sample data is provided in Appendix~\ref{sec:Samples of BlendNet}.

\subsection{Model Optimization}
The development of BlenderLLM involves a two-phase optimization process: Supervised Fine-tuning (SFT) and Self-improvement.

\subsubsection{Step I: Supervised Fine-tuning}

We utilize the aforementioned data to fine-tune the \texttt{Qwen2.5-Coder-7B-Instruct} model, thereby obtaining the BlenderLLM-base, which serves as the base model for the next step's optimization, denoted as \(M_0\).

\subsubsection{Step II: Self-improvement}

Due to the limited data, we employed a self-improvement approach, allowing the model to further optimize itself using data it generates. Specifically, we trained a filter with previous data to select high-quality data generated by the model, and then iteratively optimized the model through a cycle of data generation and model training.

\paragraph{Cascade Filter}
We utilize \texttt{BlendNet-Human} and \texttt{BlendNet-GPT} as positive examples. $8k$ samples are selected as negative examples from the remaining \(\langle l_j, s_j \rangle\) pairs. These data are employed to fine-tune the \texttt{Qwen2-VL-7B-Instruct} model, resulting in the Coarse Filter. Combined with \texttt{GPT-4o}, which functions as the Fine Filter, they form a Cascade Filter through a cascaded mechanism. Appendix~\ref{app:cascade_filter} summarizes the precision of each filter.

\textbf{Data Generation:}
In the \(i\)-th iteration, we generate training data using the model from the previous iteration \( M_{i-1} \). Specifically, for each instruction \( l_j \), we obtain a script \( s_j \) through inference with \( M_{i-1} \). We denote the generated dataset for iteration \( i \) as \( D_i = \{\langle l_j, s_j \rangle_i\} \). These pairs are rigorously filtered using the Cascade Filter \( F(l_j, s_j) \to \{0, 1\} \) to ensure high-quality data selection, retaining only those pairs for which \( F(l_j, s_j) = 1 \).

\textbf{Model Training:}
The selected high-quality data from the data generation phase is used to fine-tune the  model \( M_{i-1} \). This process uses the filtered data to update \( M_{i-1} \), thereby resulting \( M_{i} \). 

The process alternates between data generation and model training, creating an iterative approach to model refinement through Self-improvement, until the loss doesn't drop on the validation set. More details can be found in Appendix~\ref{sec:self_improvement}.

\section{Benchmarking CAD}
In response to the lack of a benchmark for assessing CAD script generation, we develop \texttt{CADBench}, a system designed to quantitatively evaluate this capability utilizing the method of MLLM-as-a-judge~\citep{ge2024mllmbench}. \texttt{CADBench} comprises 700 meticulously designed instructions, offering a comprehensive dataset for evaluation. Given the open-ended nature of the task, no fixed ground truth is established. Instead, the evaluation process employs a flexible and realistic framework that make the evaluation through predefined criteria.

\subsection{Design Principles}
\texttt{CADBench} is developed by the principles of user-centric, comprehensiveness, granularity and reliability.

\paragraph{User-Centric}
To simultaneously meet the diversity of test cases and align with practical applications, we constructed \texttt{CADBench-Sim} and \texttt{CADBench-Wild} through synthesized data and the collection of real data, respectively. \texttt{CADBench-Sim} provides controlled synthetic data for baseline testing, covering multiple scenarios, while \texttt{CADBench-Wild} offers real-world internet-sourced data to assess the model's practical performance and adaptability.

\paragraph{Comprehensiveness}
The comprehensive nature of \texttt{CADBench} is driven by the necessity to rigorously evaluate 3D generative models across a wide array of object categories, instruction types, and complexities. By systematically covering all categories defined in Appendices~\ref{sec:Categories, Instruction Types and Instruction Length}, the benchmark provides a robust and inclusive assessment of model performance and generalizability.

\begin{figure}
    \centering
    \includegraphics[width=0.6\linewidth]{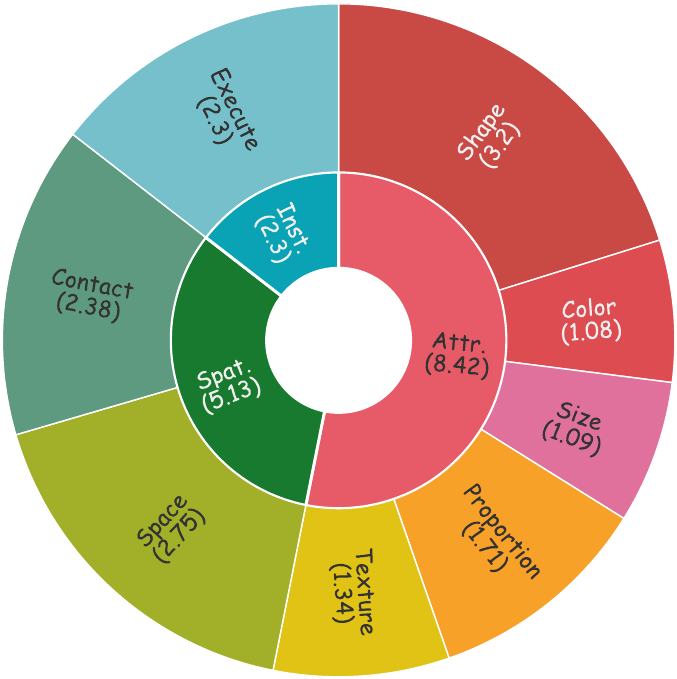}
    \caption{Dimensions of Criteria. Numbers represent the average count of criteria in that dimension.
}
    \label{fig:Dimensions of Criteria}
    
\end{figure}

\paragraph{Granularity}
The fine-grained evaluation approach of \texttt{CADBench} significantly enhances the benchmark's ability to provide detailed insights into model performance. By incorporating evaluation criteria across three dimensions, as show in Figure~\ref{fig:Dimensions of Criteria}\texttt{CADBench} ensures that models are thoroughly evaluated on diverse aspects, leading to a deeper understanding of their strengths and weaknesses. Detailed explanations and examples of each evaluation dimension are available in Appendix~\ref{sec:benchmark}.

\paragraph{Reliability}
Ensuring the reliability of \texttt{CADBench} is paramount, and this is achieved through manual annotation of grading criteria for each sample in \texttt{CADBench}. It is also  ensured by consistent evaluation and alignment with human preferences. This meticulous approach provides a dependable framework for assessing model performance, fostering trust in the results. For detailed insights into the annotation process, please refer to Appendix~\ref{sec:annotationOfCriteria}.

\subsection{CADBench Construction}

\subsubsection{Part I: CADBench-Sim}
\texttt{CADBench-Sim} comprises 500 synthetic samples. To ensure the comprehensiveness of \texttt{CADBench-Sim}, we employed the \textbf{Text Module} from Section~\ref{sec:data_construction} to generate the instruction data for \texttt{CADBench-Sim}. The resulting distribution is shown in Figure~\ref{fig:diversity_data}. 

\subsubsection{Part II: CADBench-Wild}
\texttt{CADBench-Wild} incorporates 200 real-world 3D modeling questions, sourced from various CAD-related online forums\footnote{\url{https://blenderartists.org/c/general-forums/5}\\\url{https://www.reddit.com/r/blender/}\\\url{https://discord.com/channels/185590609631903755/1006638436255551620}}. These questions represent complex, real-world scenarios that are substantially more challenging than synthetic tasks, positioning them as out-of-distribution (OOD) data relative to the training data of BlenderLLM. By reflecting actual user requirements, \texttt{CADBench-Wild} offers a critical opportunity to evaluate the generalization capacity of BlenderLLM beyond synthetic environments. The integration of these tasks ensures that \texttt{CADBench} encompasses both synthetic scenarios and real-world applications, providing a comprehensive assessment for the LLMs.

\subsection{Criteria}
Given the open-ended evaluation characteristics of CAD model assessment, we assist GPT-4o in evaluation by providing customized criteria, instead of ground truth, for each test sample. To achieve a comprehensive and detailed assessment, we designed the criteria from top to bottom into 3 major dimensions and 8 minor dimensions, as shown in the Figure~\ref{fig:Dimensions of Criteria}. After determining the criteria dimensions, we employ GPT-4o to generate a draft criteria for each sample, and thenmanually verify the criteria following the instruction in Appendix~\ref{sec:annotationOfCriteria}, with criteria examples available in the Appendix~\ref{app:criteria_example}. The introduction of criteria not only enhances the comprehensiveness of the evaluation but also improves the consistency between model assessment and human evaluation, as mentioned in the next section.

\subsection{Evaluation Protocol}

\paragraph{Evaluation Procedure} \texttt{CADBench} operates through three distinct stages.

The first stage is script generation. Let \( e \) represent the one-shot example used to guide the LLM. The LLM generates a \texttt{bpy} script \( s = f(l, e)\) based on these instructions and the context. This ensures improved responses and maintains comparability with BlenderLLM's results.

Second, the generated script \( s \) is executed in Blender to produce a set of rendered images \( I = \{i_1, i_2, i_3, i_4\} \), where each \( i_k \) is a screenshot captured from different angles.

Finally, these images \( I \) along with the script are evaluated by \texttt{GPT-4o} using predefined scoring criteria. For each criterion \( c_i \), we define the evaluation function \( E(l, I, s, c_i) \to \{0, 1\} \), where \( E(l, I, s, c_i) = 1 \) if the criterion is satisfied and \( 0 \) otherwise.

\paragraph{Evaluation Methodology} 
To accurately assess the generated CAD outputs from different aspects, we employ \texttt{GPT-4o} for two complementary evaluation approaches:

\begin{itemize}
    \item \textbf{Image-Based Evaluation:} This approach targets the spatial aspects of the CAD scripts which are hard to evaluate without image. Each criterion \( c_i  \) is assessed for visual fidelity using the evaluation function \( E_I(l, I, c_i) \).

    \item \textbf{Script-Based Evaluation:} To accurately assess objective attributes such as size, color, and material, which are challenging to evaluate visually, we evaluate directly using the \texttt{bpy} script \( s \). The evaluation function \( E_S(l, s, c_i) \) ensures precise scoring of these attributes.
\end{itemize}

The detailed evaluation process is provided in Appendix~\ref{sec:eval_process}.

\paragraph{Evaluation Reliability} To verify the reliability of the LLM-as-a-Judge framework, two human evaluators independently review a sample of 200 outputs from different models. Appendix~\ref{sec:annotationOfScoring} presents the details of the manual annotation for evaluation. And the human evaluation resulted in a \textit{kappa} value of 0.883. The inter-rater reliability between LLM and the human evaluators is calculated using Cohen’s kappa coefficient, yielding a \textit{kappa} value of 0.791, which signifies a high level of agreement.

\subsection{Evaluation Metrics}

 For each model, the final score is calculated by averaging the outputs across all criteria: 

\[
Score = \frac{1}{|C|} \sum_{c_i \in C} E(l, I, s, c_i)
\]

Note that for some of the criteria, the image input \( I \) is empty, while for others, script input \( s \) is empty. See Appendix~\ref{sec:metrics} for more details.

\begin{table*}[]
\centering
\resizebox{\textwidth}{!}{%
\begin{tabular}{l|ccccc|ccccc}
\toprule
\multirow{2}{*}{\textbf{Models}} &
  \multicolumn{5}{c|}{\textbf{CADBench-Sim}} &
  \multicolumn{5}{c}{\textbf{CADBench-Wild}} \\ \cmidrule(lr){2-11}
 &
  \textit{Attr.}$\uparrow$ &
  \textit{Spat.}$\uparrow$ &
  \textit{Inst.}$\uparrow$ &
  \textit{Avg.}$\uparrow$ &
  $E_{syntax}$$\downarrow$ &
  \textit{Attr.}$\uparrow$ &
  \textit{Spat.}$\uparrow$ &
  \textit{Inst.}$\uparrow$ &
  \textit{Avg.}$\uparrow$ &
  $E_{syntax}$$\downarrow$ \\ 
    \midrule
    \rowcolor{mygray}\multicolumn{11}{c}{\textit{Closed-source Models}}\\
\textbf{o1-Preview}        & 0.729 & 0.707 & 0.624 & $0.687 \pm 0.045$ & 15.6\% & 0.595 & 0.612 & 0.542 & $0.583 \pm 0.030$ & 17.5\% \\ 
\textbf{GPT-4-Turbo}       & 0.658 & 0.621 & 0.488 & $0.589 \pm 0.073$ & 18.2\% & 0.526 & 0.541 & 0.478 & $0.515 \pm 0.027$ & 24.5\% \\
\textbf{Claude-3.5-Sonnet} & 0.687 & 0.608 & 0.482 & $0.593 \pm 0.084$ & 15.6\% & 0.529 & 0.508 & 0.43 & $0.489 \pm 0.043$ & 26.5\% \\
\textbf{GPT-4o}           & 0.623 & 0.593 & 0.479 & $0.565 \pm 0.062$ & 21.4\% & 0.460 & 0.466 & 0.408 & $0.444 \pm 0.026$ & 28.5\% \\
\textbf{BlenderGPT}        & 0.574 & 0.540 & 0.444 & $0.519 \pm 0.055$ & 25.2\% & 0.402 & 0.425 & 0.368 & $0.398 \pm 0.023$ & 35.0\% \\ 
\textbf{Gemini-1.5-Pro}    & 0.535 & 0.483 & 0.387 & $0.468 \pm 0.061$ & 30.2\% & 0.375 & 0.404 & 0.361 & $0.380 \pm 0.018$ & 38.0\% \\
    \midrule
    \rowcolor{mygray}\multicolumn{11}{c}{\textit{Open-source Models}}\\
\textbf{DeepSeek-V2.5}       & 0.569 & 0.497 & 0.372 & $0.479 \pm 0.081$ & 25.2\% & 0.422 & 0.394 & 0.345 & $0.387 \pm 0.032$ & 34.0\% \\
\textbf{Qwen2.5-Coder-7B-Instruct}      & 0.457 & 0.352 & 0.251 & $0.353 \pm 0.084$ & 31.4\% & 0.354 & 0.327 & 0.250 & $0.310 \pm 0.044$ & 37.0\% \\
\textbf{Qwen2.5}          & 0.367 & 0.274 & 0.193 & $0.278 \pm 0.071$ & 44.8\% & 0.220 & 0.219 & 0.170 & $0.203 \pm 0.023$ & 58.5\% \\
\textbf{LLaMA-3.1-8B-Instruct} & 0.125 & 0.087 & 0.071 & $0.094 \pm 0.023$ & 76.0\% & 0.130 & 0.127 & 0.105 & $0.120 \pm 0.011$ & 65.5\% \\
\textbf{Mistral-7B-Instruct-V0.3}      & 0.015 & 0.018 & 0.015 & $0.016 \pm \textbf{0.001}$ & 96.8\% & 0.023 & 0.031 & 0.030 & $0.028 \pm \textbf{0.004}$ & 93.0\% \\
\textbf{CodeLLaMA-7B-Instruct}      & 0.005 & 0.004 & 0 & $0.003 \pm 0.002$ & 98.8\% & 0.009 & 0.019 & 0.015 & $0.014 \pm \textbf{0.004}$ & 96.5\%\\
    \midrule
    \rowcolor{mygray}\multicolumn{11}{c}{\textit{BlenderLLMs (Ours)}}\\
\textbf{Iteration 1}     & 0.784 & 0.689 & 0.517 & $0.663 \pm 0.111$ & 5.8\% & 0.673 & 0.569 & 0.444 & $0.562 \pm 0.094$ & 6.0\% \\
\textbf{Iteration 2}     & 0.822 & 0.743 & 0.597 & $0.721 \pm 0.093$ & 5.2\% & 0.689 & 0.608 & 0.473 & $0.590 \pm 0.089$ & 6.0\% \\
\textbf{Iteration 3}     & \textbf{0.846} & 0.760 & \textbf{0.638} & $\textbf{0.748} \pm 0.085$ & 3.4\% & \textbf{0.739} & \textbf{0.675} & \textbf{0.578} & $\textbf{0.664} \pm 0.066$ & \textbf{3.5\%} \\
\textbf{Iteration 4}     & \textbf{0.846} & \textbf{0.767} & 0.626 & $0.747 \pm 0.091$ & \textbf{3.2\%} & 0.717 & 0.614 & 0.493 & $0.608 \pm 0.092$ & 5.0\% \\ \bottomrule
\end{tabular}%
}
\caption{Quantitative Assessment for Instruction-to-Script Generation. This table compares the performance of 12 LLMs and BlenderLLM in assisting CAD script generation on \texttt{CADBench} across three dimensions: \textit{Attr.}, \textit{Spat.}, and \textit{Inst.}. Additionally, \textit{Avg.} and \(E_{syntax}\) are provided. A higher score indicates better performance in a given dimension. The results show that BlenderLLM outperforms all other models and effectively handles the task of Instruction-to-CAD script generation.
}
\label{tab:scores}
\end{table*}

\section{Experiments}

\subsection{Training Details}

We use \texttt{Qwen2.5-Coder-7B-Instruct} as the base model and fine-tune it on \texttt{BlendNet-Human} to obtain the BlenderLLM-base. For subsequent rounds, the input data size is fixed at $2k$ samples to prevent training data saturation and overfitting. During the SFT, full parameter fine-tuning is applied. Each model training session is conducted on four A800 GPUs with 80GB of memory, with a training time of approximately 21 minutes per SFT round. The batch size, gradient steps, learning rate, epochs, and warmup ratio are set to 1, 2, \(1 \times 10^{-5}\), 1, and 0.1, respectively. The validation dataset constitutes 10\% of the total dataset, with a batch size of 1 and 50 evaluation steps.

\subsection{Baselines}
\label{sec:baseline}

To evaluate the performance of BlenderLLM, we compare it against several existing models using a one-shot context approach for all comparisons. The models used for comparison include:\textbf{o1-Preview} \citep{o1preview2024}, \textbf{GPT-4 turbo} \citep{openai2023gpt4}, \textbf{Claude3.5-sonnet} \citep{anthropic2023claude},  \textbf{GPT-4o} \citep{openai2024gpt4o}, \textbf{BlenderGPT} \citep{blendergpt2023},  \textbf{Gemini-1.5-pro} \citep{googlegemini2024}, \textbf{DeepSeek-V2.5} \citep{deepseek2024}, \textbf{Qwen2.5-Coder-7B-Instruct} \citep{qwen25_2024}, \textbf{Qwen-2.5} \citep{qwen25_2024}, \textbf{LLaMA3.1-8B-Instruct} \citep{touvron2023llama}, \textbf{Mistral-7B-Instruct-V0.3} \citep{mistral_2024}, and \textbf{CodeLLaMa-7B-Instruct} \citep{codellama2024}. Details about these models can be found in Appendix~\ref{sec:baselines}. 

\subsection{Main Results}
\label{sec:results}

\paragraph{Overall Performance}
As shown in Table~\ref{tab:scores}, BlenderLLM achieves SOTA performance across all dimensions in both \texttt{CADBench-Sim} and \texttt{CADBench-Wild}, significantly outperforming the second-place model, o1-Preview. A visual comparison of the performance of different models across the dimensions of \textit{attr.}, \textit{spat.}, and \textit{inst.} is provided in Appendix~\ref{sec:The Visual Examples of the Performance of Different Models}, where it is evident that BlenderLLM demonstrates substantial improvements in all three dimensions. Furthermore, the comparison shows that BlenderLLM not only adheres more closely to the specified requirements but also offers more reasonable solutions for unmentioned aspects. Its strong performance on \texttt{CADBench-Wild} further highlights BlenderLLM's exceptional generalization capabilities.

\begin{table}[h!]
    \centering
    \renewcommand{\arraystretch}{1} 
    \begin{tabular}{
        |>{\centering\arraybackslash}m{2cm}|
        >{\centering\arraybackslash}m{4.5cm}|
    }
        \hline
        \multicolumn{2}{|p{6.5cm}|}{\small \textbf{Instruction:} \textit{Create a desktop monitor. It should have a 24-inch screen with a thin bezel.}} \\ \hline
        \small \textbf{Iteration} & \small \textbf{Images} \\ \hline
        \small Base Model & \includegraphics[width=4cm]{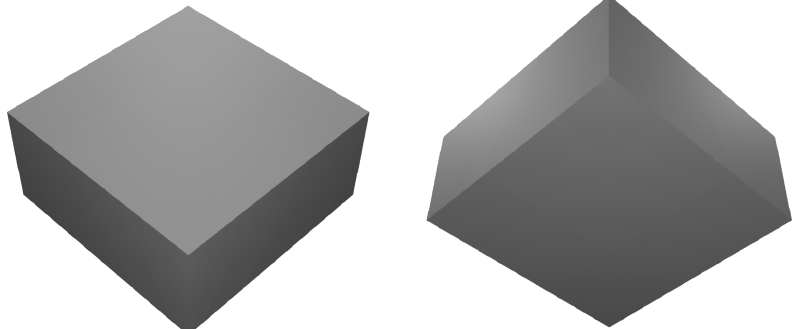} \\ \hline
        \small Iteration 1 & \includegraphics[width=4cm]{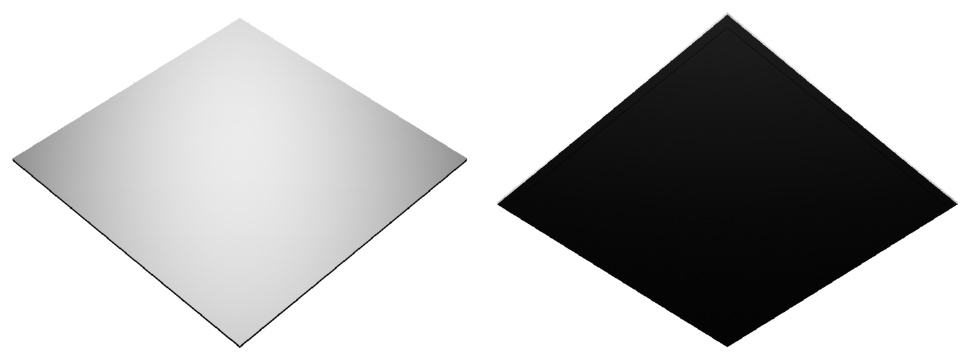} \\ \hline
        \small Iteration 2 & \includegraphics[width=4cm]{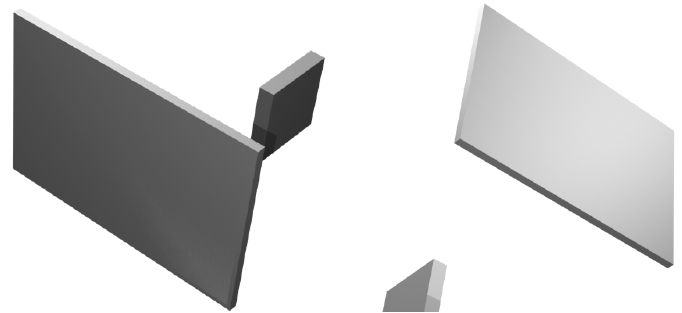} \\ \hline
        \small Iteration 3 & \includegraphics[width=4cm]{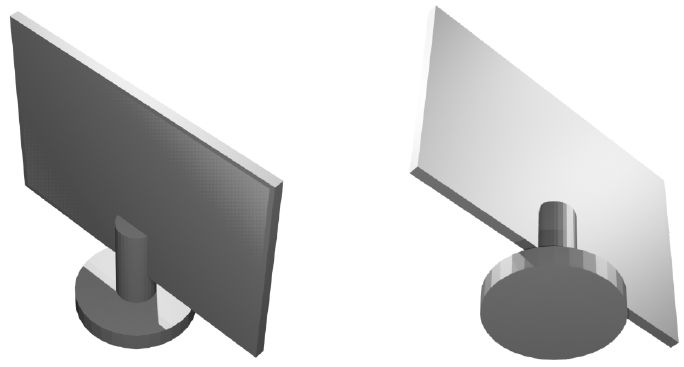} \\ \hline
        \small Iteration 4 & \includegraphics[width=4cm]{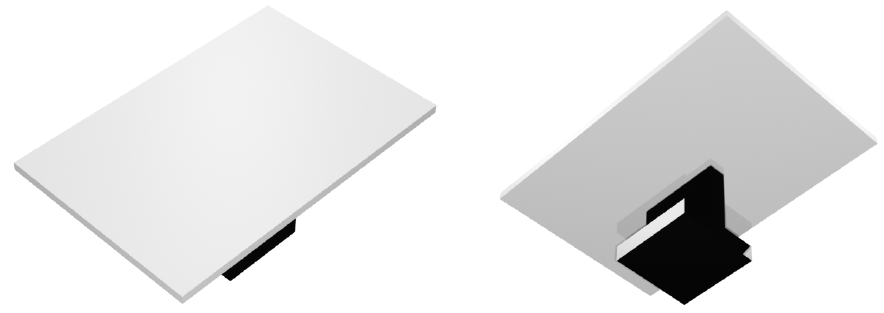} \\ \hline
    \end{tabular}
    \caption{Visual Process of Self-improvement}
    \label{tab:self-improvement}
\end{table}

\paragraph{Syntax Error Rate}
As BlenderLLM fine-tuned with high-quality specialized data, its syntax error rate is significantly lower than that of other models. Moreover, the syntax error rate on \texttt{CADBench-Wild} has barely increased, further demonstrating that BlenderLLM has achieved a high level of proficiency in understanding CAD script syntax.

\paragraph{Self-improvement}
As shown in the examples in Table~\ref{tab:self-improvement}, during the Self-improvement process, BlenderLLM evolves from initially having limited ability to follow instructions, to gradually understanding the instructions and developing spatial reasoning capabilities, ultimately succeeding in modeling the specified object.

\subsection{Analysis and Discussion}
The experimental results demonstrate that BlenderLLM exhibits significant advantages in \textit{attr.}, \textit{spat.}, \textit{inst.}, and $E_{syntax}$. Combining the performance of different models on sub-dimensions, as shown in Appendix~\ref{sec:Performance of LLMs on Sub-Dimension}, with the comparison of visualization results presented in Appendix~\ref{sec:The Visual Examples of the Performance of Different Models} and Table~\ref{sec:self_improvement}, these achievements can be attributed to two key factors. First, the \texttt{BlendNet} enables BlenderLLM to learn a variety of instructions. Also, This comprehensive training helped BlenderLLM develop a deeper understanding of the rationality of object attributes, such as the relative size and position of components, as well as the matching of colors and materials. Second, the Self-improvement training strategy allowed BlenderLLM to continuously learn and adapt, progressively enhancing its spatial reasoning capabilities over iteration.

\section{Ablation}

\label{sec:ablation}
\begin{table}[t]
    \centering
    \resizebox{\linewidth}{!}{%
        \begin{tabular}{l|cc|cc}
        \toprule
            \multirow{2}{*}{\textbf{Methods}} &  \multicolumn{2}{c|}{\textbf{CADBench-Sim}} & \multicolumn{2}{c}{\textbf{CADBench-Wild}}\\
            & \textit{Avg.} & $E_{\text{syntax}}$ & \textit{Avg.} & $E_{\text{syntax}}$ \\
                \midrule
                \rowcolor{mygray}\multicolumn{5}{c}{\textit{Epoch Accumulation Training}}\\
            \textit{+ 1 epoch}  & $0.663 \pm 0.111$ & 5.8\% & $0.562 \pm 0.094$ & 6.0\% \\
            \textit{+ 2 epoch}  & $0.685 \pm 0.105$ & 5.6\% & $0.578 \pm 0.086$ & 5.0\% \\
            \textit{+ 3 epoch}  & $0.721 \pm 0.099$ & 3.6\% & $0.568 \pm 0.089$ & 6.5\% \\
            \textit{+ 4 epoch}  & $0.705 \pm 0.103$ & \textbf{3.2\%} & $0.595 \pm 0.082$ & 6.0\% \\

                \midrule
                \rowcolor{mygray}\multicolumn{5}{c}{\textit{Predefined Incremental Training}}\\
            \textit{+ 1 increment}  & $0.663 \pm 0.111$ & 5.8\% & $0.562 \pm 0.094$ & 6.0\% \\
            \textit{+ 2 increment}  & $0.716 \pm 0.098$ & 4.8\% & $0.559 \pm 0.088$ & 5.5\% \\
            \textit{+ 3 increment}  & $0.722 \pm 0.099$ & 3.6\% & $0.593 \pm 0.080$ & 6.5\% \\
            \textit{+ 4 increment}  & $0.721 \pm 0.098$ & 3.8\% & $0.606 \pm 0.087$ & 5.0\% \\

                \midrule
                \rowcolor{mygray}\multicolumn{5}{c}{\textit{Self-improvement Training}}\\
            \textit{+ 1 iteration}  & $0.663 \pm 0.111$ & 5.8\% & $0.562 \pm 0.094$ & 6.0\% \\
            \textit{+ 2 iteration}  & $0.721 \pm 0.093$ & 5.2\% & $0.590 \pm 0.089$ & 6.0\% \\
            \textit{+ 3 iteration}  & $\textbf{0.748} \pm \textbf{0.085}$ & 3.4\% & $\textbf{0.664} \pm \textbf{0.066}$ & \textbf{3.5\%} \\
            \textit{+ 4 iteration}  & $0.747 \pm 0.091$ & \textbf{3.2\%} & $0.608 \pm 0.092$ & 5.0\% \\
            \bottomrule
        \end{tabular}%
    }
    \caption{Comparison between different SFT strategy.}
    \label{tab:sft_strategy}
\end{table}

To demonstrate that Self-improvement Training strategy is more effective than conventional iterative training strategy with similar computational resources, we conducte two comparative experiments:

\paragraph{Epoch Accumulation Training}
We fine-tune \texttt{Qwen2.5-Coder-7B-Instruct}, using the fixed dataset \texttt{BlendNet-Human}. The training process begin with one epoch and is incrementally extended by adding an additional epoch in each iteration.

\paragraph{Predefined Incremental Training}
We fine-tuned the base model, \texttt{Qwen2.5-Coder-7B-Instruct}, using a predefined incremental strategy. The process began with the initial dataset, \texttt{BlendNet-Human}. In subsequent iterations, $2k$ unused examples from \texttt{BlendNet-GPT} were added for further fine-tuning.

Table~\ref{tab:sft_strategy} demonstrates that, after the same number of training iterations, models trained using the Self-improvement Training strategy consistently outperform those trained with the other two approaches on both \texttt{CADBench-Sim} and \texttt{CADBench-Wild}. Furthermore, Appendix~\ref{sec:The Visual Examples of the Performance of Different Training strategy} presents the visualization results of the three different training strategies. It can be observed that, compared to the other two strategies, the Self-improvement Training strategy exhibits superior performance in both instruction-following and spatial reasoning capabilities.

\section{Conclusion}

In this paper, we propose a comprehensive framework that spans from data construction to self-improvement-based SFT model training and benchmark testing. Through this framework, BlenderLLM, has demonstrated superior performance across various metrics compared to mainstream models. Our results highlight the effectiveness of combining Self-improvement with high-quality dataset, leading to significant advancements in model capabilities.

\newpage

\section*{Limitation}

This study has several limitations. First, the data construction and model training primarily focused on basic CAD modeling aspects and did not address more intricate elements, such as material properties, surface treatments, or internal complexity. These factors could influence the model’s performance in handling more advanced CAD tasks. Second, our work focused solely on generating CAD scripts from user instructions, without exploring the potential for direct CAD model generation or the integration of multimodal inputs, such as combining user instructions with images. Future research could investigate these avenues to enhance model versatility. Lastly, the model has not been trained for multi-turn dialogues, limiting its ability to engage in more complex, interactive conversations. These limitations highlight key areas for future improvement and expansion of the model’s capabilities.

\section*{Ethics Statement}
This research involves the development and evaluation of a novel dataset and methodology for applying Large Language Models (LLMs) to Computer-Aided Design (CAD). The study does not involve human subjects, nor does it utilize any personally identifiable information. The research adhere to ethical guidelines regarding data privacy and intellectual property.  The authors declare no conflicts of interest related to this work. The datasets and models we provide follow the CC-BY 4.0 License. 

\newpage

\bibliography{custom}

\clearpage
\appendix

\label{sec:appendix}

\section{Comparison of BlenderLLM and Recent works}
\label{sec:Comparison of BlenderLLM and Recent Works}

The comparison of BlenderLLM and recent works is shown in Table~\ref{tab:LLMs4CAD}.

\begin{table*}[t]
\centering
\resizebox{\textwidth}{!}{%
\fontsize{10}{14}\selectfont
\begin{tabular}{l|cccccc}
\hline
\textbf{Models} &
  \textbf{\begin{tabular}[c]{@{}c@{}}Open Source\end{tabular}} &
  \textbf{\begin{tabular}[c]{@{}c@{}}Self-improvement\end{tabular}} &
  \textbf{\begin{tabular}[c]{@{}c@{}}Methodology \end{tabular}} &
  \textbf{\begin{tabular}[c]{@{}c@{}}LM Backbone\end{tabular}} &
  \textbf{Size} &
  \textbf{Task} \\ \hline
{BlenderGPT \cite{blendergpt2023}} & \ding{55}  & \ding{55}  & Prompt Engineering & GPT-4 \cite{openai2023gpt4} & /       & Text-to-Code  \\
{CADGPT \cite{kapsalis2024cadgptharnessingnaturallanguage}}     & \ding{55}  & \ding{55}  & Prompt Engineering & GPT-4 \cite{openai2023gpt4} & /       & Text-to-API  \\
{CAD-LLM \cite{wu2023cadllm}}    & \ding{55}  & \ding{55}  & Training           & T5 \cite{t5}    & 770M    & CAD-to-CAD \\
{CADVLM \cite{CadVLM2024}}     & \ding{55}  & \ding{55}  & Training & /  & /       & Multimodal-to-CAD \\ \hline
\textbf{BlenderLLM} & \ding{51} & \ding{51} & Training           & Qwen2.5-Coder \cite{qwen25_2024} & 7B      & Text-to-Code \\ \hline
\end{tabular}%
}
\caption{Comparison of BlenderLLM and Recent Works}
\label{tab:LLMs4CAD}
\end{table*}

\section{Data Construction}
\subsection{Categories, Instruction Types and Instruction Length}
\label{sec:Categories, Instruction Types and Instruction Length}

\subsubsection{Categories}
\label{Category List}

We based on the Locarno Classification System to generate our own classification method and concluded all objects into 16 categories \( C = \{\textit{Tech}, \textit{Music}, \ldots, \textit{Home}\} \), with their names listed below:
\begin{itemize}
    \item \textit{Tech}: Recording, telecommunication, or data processing equipment
    \item \textit{Music}: Musical instruments
    \item \textit{Animal}: Articles for the care and handling of animals
    \item \textit{Furn}: Furnishing
    \item \textit{Transport}: Means of transport or hoisting
    \item \textit{Office}: Stationery and office equipment, artists' and teaching materials
    \item \textit{Food}: Foodstuffs
    \item \textit{MedLab}: Medical and laboratory equipment
    \item \textit{Fashion}: Articles of clothing and haberdashery
    \item \textit{Graphics}: Graphic symbols, logos, surface patterns, ornamentation, arrangement of interiors and exteriors
    \item \textit{Recre}: Recreational goods (Games, toys, tents, and sports goods)
    \item \textit{Tools}: Tools and hardware
    \item \textit{Travel}: Travel goods, cases, parasols, and personal belongings, not elsewhere specified
    \item \textit{Power}: Electrical systems (Equipment for production, distribution, or transformation of electricity)
    \item \textit{Cuisine}: Culinary machines (Machines and appliances for preparing food or drink, not elsewhere specified)
    \item \textit{Home}: Household goods, not elsewhere specified
\end{itemize}

\subsubsection{Instruction Types}
\label{Instruction Type}

We notice the difference between styles of prompting. In order to make input data more diverse, we specified them into 8 types, denoted as \( T = \{\textit{Verbal}, \textit{Look}, \ldots, \textit{Design}\} \), with their names listed below:
\begin{itemize}
    \item \textit{Verbal}: Verbal Question\\
    Direct and conversational requests for creating dynamic or specific action images, focusing on movement and behavior.    
    \item \textit{Look}: Outlook Question\\
    Focuses on the physical appearance of objects, emphasizing visual attributes like color and shape.    
    \item \textit{Use}: Specific Usage Question\\
    Emphasizes the practicality or functionality of objects, highlighting how they can be used or their intended purpose.    
    \item \textit{Deco}: Decoration Question\\
    Concentrates on the aesthetic or decorative aspects of objects, underlining their decorative value and appearance.
    \item \textit{Feel}: Feeling Question\\
    Involves sensory experiences or the tactile quality of objects, aiming to capture the feel or sensory impression they convey.
    \item \textit{Comp}: Comparing Question\\
    Entails making distinctions based on comparison, often with a focus on historical or time-specific characteristics to capture a specific style.
    \item \textit{Feat}: Feature Question\\
    Centers around exploring and describing specific features of objects, requiring creativity based on given characteristics.
    \item \textit{Design}: Design Question\\
    Revolves around creative construction or conceptualization based on specific shapes or ideas, emphasizing innovative design solutions.
\end{itemize}

\subsubsection{Instruction Length}
\label{length list}

We set the length of the instruction to enhance the variety. We place instruction into 5 classes regarding to their words count, as \( L = \{\textit{VS}, \textit{S}, \ldots, \textit{E}\}\).
\begin{itemize}
    \item \textit{VS}: Very Short
    \item \textit{S}: Short
    \item \textit{M}: Medium
    \item \textit{L}: Long
    \item \textit{E}: Extended
\end{itemize}

\subsection{Instruction Generation Process}
\label{sec:Instruction Generation Process}

the generation process for instructions is shown in Algorithm~\ref{Algorithm:Iterative Generation and Normalization Process}

\begin{algorithm}[h]
\small
\caption{Instruction Generation Process}
\label{Algorithm:Iterative Generation and Normalization Process}
\begin{algorithmic}[1]
\State \textbf{Input:}
\State \quad $I$: Set of instructions
\State \quad $I_{\text{prev}}$: Set of previous instructions
\State \quad $I_{\text{seed}}$: Set of seed instructions
\State \quad $C$: Set of categories
\State \quad $T$: Set of types
\State \quad $L$: Set of lengths
\State \quad $D_{\text{old}}$: Dataset of old instructions
\State \quad $\text{threshold}$: Threshold for name counts
\State \quad $S$: Similarity score function

\State \textbf{Output:}
\State \quad $I_{\text{new}}$: Set of new instructions
\State \quad $C_{\text{new}}$: Set of new categories
\State \quad $T_{\text{new}}$: Set of new types
\State \quad $L_{\text{new}}$: Set of new lengths
\State \quad $N_{\text{normalized}}$: Normalized names set
\State \quad $N_{\text{throwed}}$: Filtered names (names to avoid)
\State \quad $N_{\text{remaining}}$: Remaining names

\State \textbf{Iterative Generation:}
\State $I_{\text{new}} \gets \{ i \in I \mid S(i, j) < 0.8, \forall j \in I_{\text{prev}} \cup I_{\text{seed}} \}$
\State $\left| I_{\text{new}} \right| \gets 10$

\State \textbf{Constraints:}
\State $C_{\text{new}} \gets \{ C_i \mid C_i \in C, \ |C_{\text{new}}| = 16 \}$
\State $T_{\text{new}} \gets \{ T_i \mid T_i \in T, \ |T_{\text{new}}| = 8 \}$
\State $L_{\text{new}} \gets \{ L_i \mid L_i \in L, \ |L_{\text{new}}| = 5 \}$

\State \textbf{Normalization and Filtering:}
\State \textbf{1. Normalized Names Set:}
\State $N_{\text{normalized}} \gets \{ \text{normalize}(d_{\text{name}}) \mid d \in D_{\text{old}}, \ d_{\text{category}} \in C_{\text{new}} \}$

\State \textbf{2. Name Counts:}
\State $N_{\text{counts}} \gets \text{Counter} \left( \text{normalize}(d_{\text{name}}) \mid d \in D_{\text{old}} \right)$
\State \textbf{3. Filtered Names (Names to Avoid):}
\State $N_{\text{throwed}} \gets \{ n \mid n \in N_{\text{normalized}}, N_{\text{counts}}[n] > \text{threshold} \}$
\State $N_{\text{remaining}} \gets N_{\text{normalized}} - N_{\text{throwed}}$

\State \textbf{Output:}
\State \quad $I_{\text{new}}, C_{\text{new}}, T_{\text{new}}, L_{\text{new}}, N_{\text{normalized}}, N_{\text{throwed}}, N_{\text{remaining}}$
\end{algorithmic}
\end{algorithm}

\subsection{Validation}
\label{sec:validation}

\subsection{Script Generation}
\label{sec:Script Generation}
\begin{figure*}[htbp]
    \centering
    \includegraphics[width=\textwidth]{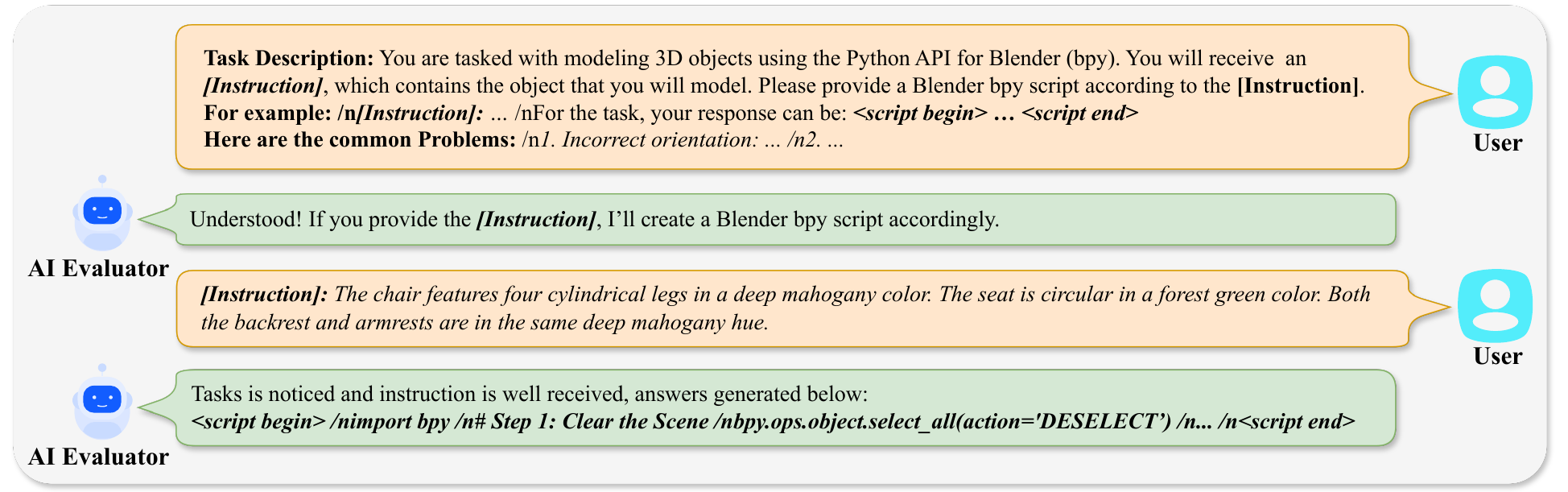}
    \caption{Process for Script Generation. We carefully designed the prompt to maximize the responsiveness and effectiveness of \texttt{GPT-4o}, ensuring that it generates high-quality and contextually accurate CAD scripts.}
    \label{fig:Prompt Script Generator}
\end{figure*}

The process for script generation is shown in Figure~\ref{fig:Prompt Script Generator}.

\subsection{Validation Process}
\label{sec:Validation Process}
\begin{figure*}[htbp]
    \centering
    \includegraphics[width=\textwidth]{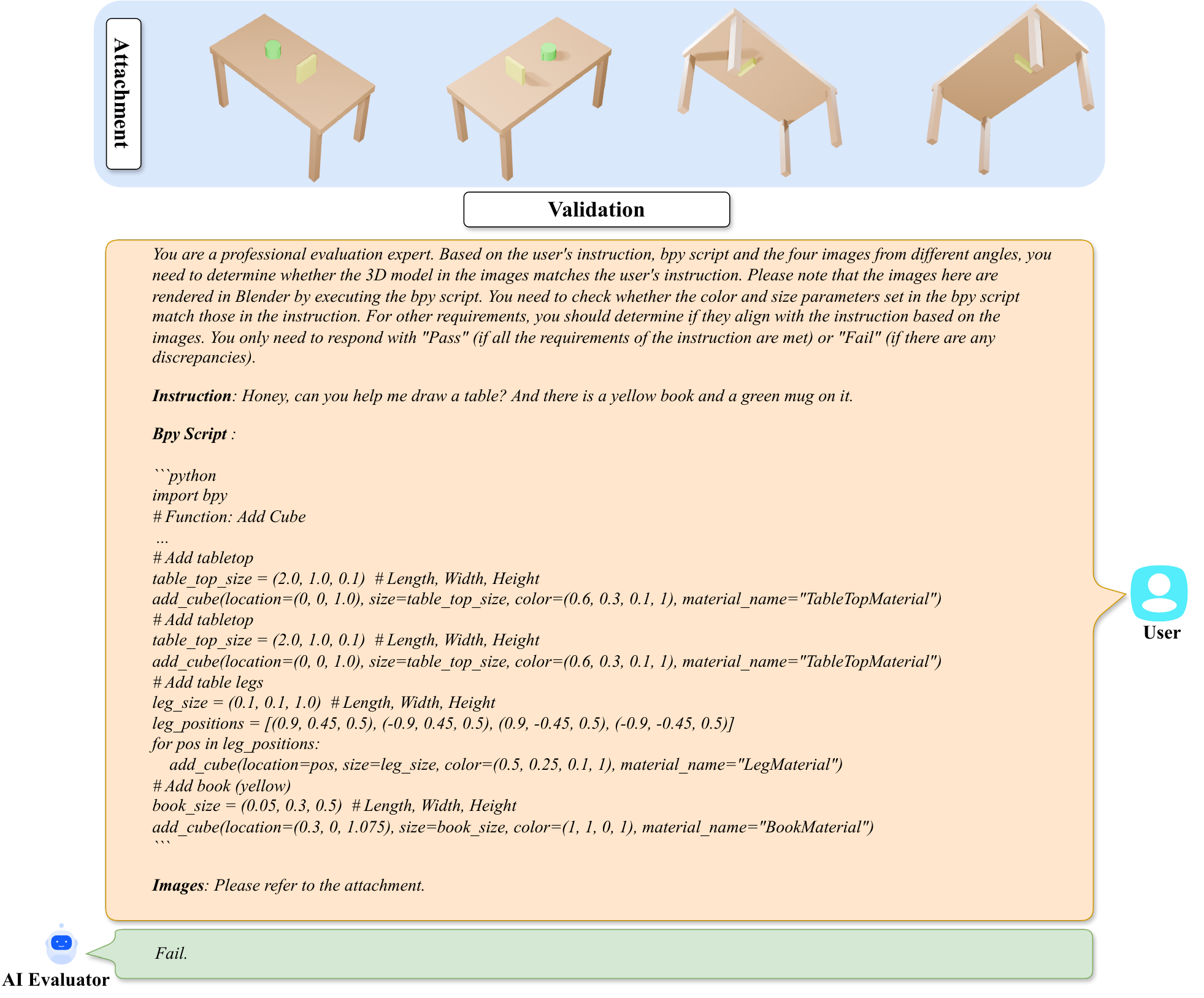}
    \caption{Validation Process}
    \label{fig:Validation Process}
\end{figure*}

The process for validation is shown in Figure~\ref{fig:Validation Process}.

\subsection{Cross Validation}
\label{sec:cross_validation}
Table~\ref{tab:cross_validation} shows the details about the cross validation result. The proportion of samples where humans and models consistently judge passed is 21.6\%, the proportion of samples where humans and models consistently judge not passed is 68.1\%, and the proportion of samples where human and model judgments differ is only 10.3\%, which demonstrates a high degree of consistency between human and model assessments. The instruction for human validation can be found in Appendix~\ref{sec:BlendNet-Human}.

\begin{table}
\centering
\fontsize{10}{10}\selectfont 
\begin{tblr}{
  row{2} = {c},
  row{3} = {c},
  cell{1}{2} = {c},
  cell{1}{3} = {c},
  hlines,
  vlines,
}
\diagbox{\textbf{GPT}}{\textbf{Human}} & \textit{Pass} & \textit{Fail}\\
\textit{Pass} & 21.61\% & 7.20\%\\
\textit{Fail} & 3.13\% & 68.06\%
\end{tblr}
\caption{Cross Validation}
\label{tab:cross_validation}
\end{table}

\subsection{The Complexity of BlendNet}
\label{sec:complexity_BlendNet}

we define three key metrics to quantify the complexity of \texttt{BlendNet}:

\begin{itemize}
    \item \textbf{Unit Number}: This metric represents the number of basic shapes within the 3D Model. It serves as an indicator of geometric complexity, where higher values imply a greater number of components and higher structural complexity.
    
    \item \textbf{Parameter Density}: This metric calculates the average complexity per shape, defined as:
    \begin{equation} \small
    Parameter\ Density = \frac{Parameter\ \#}{Unit\ \#}
    \end{equation}
    A higher parameter density indicates that each shape is more parameterized, implying greater irregularity and higher computational complexity. This value reflects how intricately the shapes are defined and how complex the relationships between the parameters are within the 3D model.
    
    \item \textbf{Entropy}: Entropy measures the spatial diversity of the shapes in the 3D space. It is defined as:
    \begin{equation}
    H = -\sum p_i \log(p_i)
    \end{equation}
    where \( p_i \) is the probability density in 3D voxels. Higher entropy values indicate greater spatial diversity, which implies more irregular and unpredictable configurations. This metric helps capture the distribution and variation of shapes across the 3D space, with larger values corresponding to more complex and diverse spatial arrangements.
\end{itemize}

\begin{figure*}[t]
    \centering
    \includegraphics[width=\textwidth]{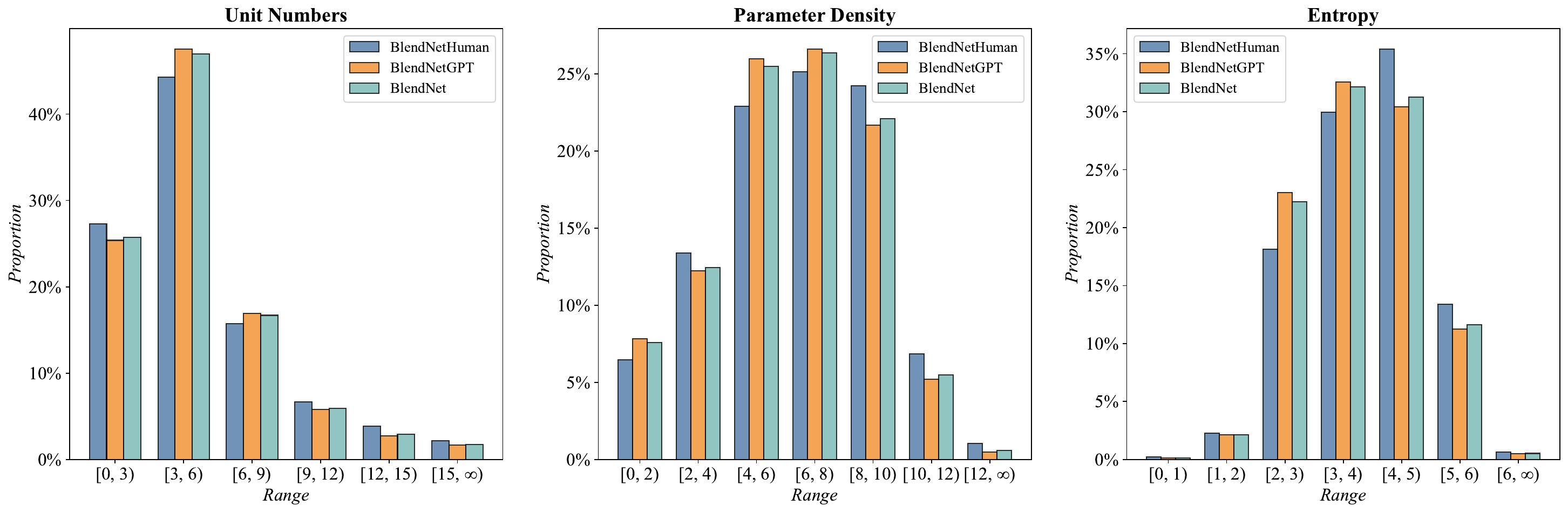}
    \caption{The complexity distribution of BlendNet}
    \label{fig:sta_blendnet}
\end{figure*}

The distribution of \texttt{BlendNet-Human}, \texttt{BlendNet-GPT}, and \texttt{BlendNet} across these three metrics is shown in Figure~\ref{fig:sta_blendnet}.

\subsection{Samples of BlendNet}
\label{sec:Samples of BlendNet}

The Samples of \texttt{BlendNet} is shown in Table~\ref{tab:Samples_BlendNet}.

\begin{table*}[h!]
    \centering
    \renewcommand{\arraystretch}{1.5} 
    \begin{tabular}{
        |>{\centering\arraybackslash}m{3.5cm}|
        >{\centering\arraybackslash}m{6cm}|
        >{\centering\arraybackslash}m{1.4cm}|
        >{\centering\arraybackslash}m{1.6cm}|
        >{\centering\arraybackslash}m{1.3cm}|
    }
        \hline
        \textbf{Instruction} & \textbf{Images} & \textbf{Unit Number} & \textbf{Parameter Density} & \textbf{Entropy} \\ \hline
        \small\textit{Design an eraser.} & \includegraphics[scale=0.5]{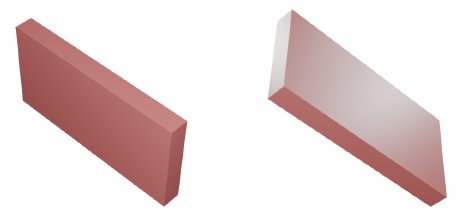} & \uline{1} & 9.00 & 2.08 \\ \hline
        \small\textit{Let's create a birthday cake with three layers. The bottom layer should be chocolate, the middle layer vanilla, and the top layer red velvet. Each layer should be separated by a thick layer of buttercream frosting. Add a decorative border of frosting around the top edge, and place colorful sprinkles all over the surface. Finally, add a Happy Birthday message on top.} & \includegraphics[scale=0.4]{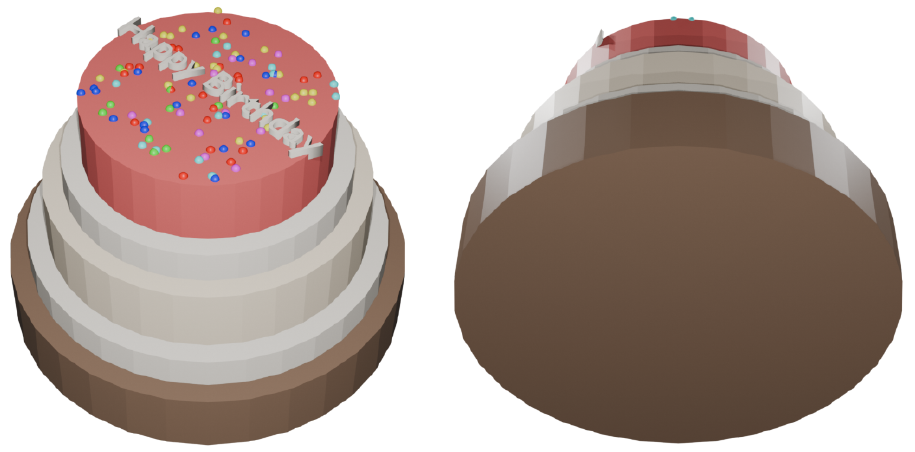} & \textbf{107} & 0.50 & 3.66 \\ \hline
        \small\textit{How does solving a puzzle cube make you feel? Can you create a 3D model of a standard 3x3 puzzle cube?} & \includegraphics[scale=0.4]{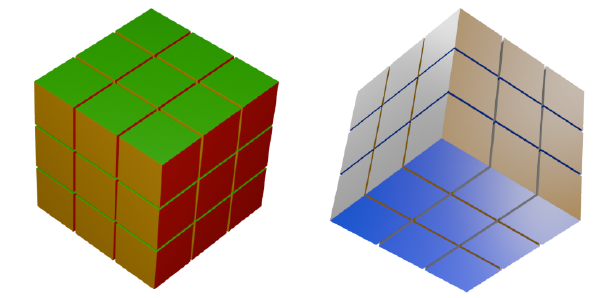} & 27 & \uline{1.41} & 3.99 \\ \hline
        \small\textit{Compare the appearance of a club sandwich and a BLT sandwich. Create both sandwiches with the classic ingredients stacked between slices of bread.} & \includegraphics[scale=0.4]{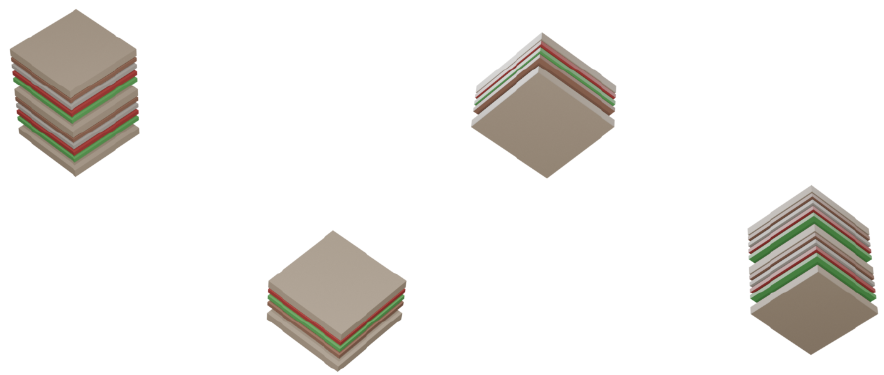} & 2 & \textbf{13.50} & 4.02 \\ \hline
        \small\textit{Design a 3D model of a smartphone with a screen and a single button on the front.} & \includegraphics[scale=0.4]{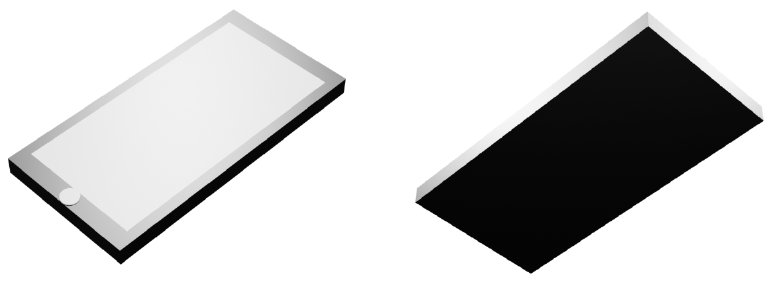} & 3 & 7.67 & \uline{1.34} \\ \hline
        \small\textit{Could you design a 3D model of a transformer coil? It should be cylindrical with multiple copper windings.} & \includegraphics[scale=0.3]{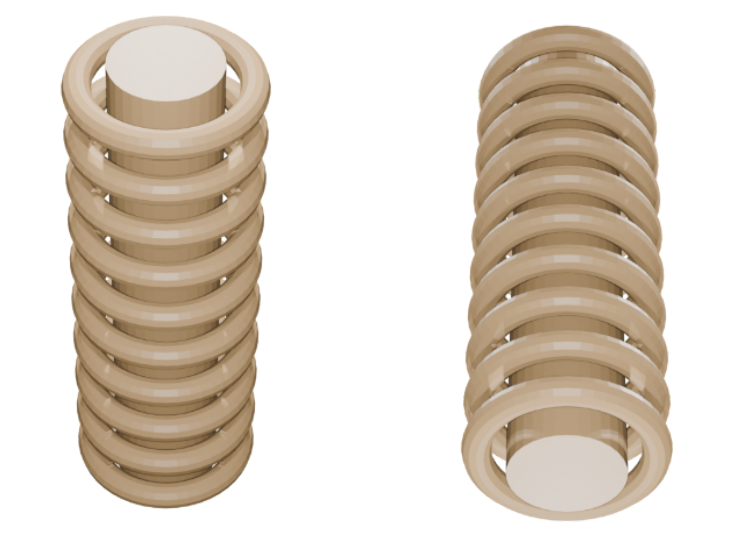} & 11 & 1.37 & \textbf{6.31} \\ \hline
    \end{tabular}
    \caption{Samples of BlendNet}
    \label{tab:Samples_BlendNet}
\end{table*}

\begin{algorithm*}[h]
\small
\caption{Self-improvement Process}
\label{algorithm:Self-improvement_process}

\begin{algorithmic}[1]
\State \textbf{Definitions:}
\State \quad $i$: Iteration number \Comment{Counter for optimization iterations, starting from 1}
\State \quad $M_i$: Model obtained at the $i$-th iteration \Comment{e.g., $M_1$ is the first iteration model}
\State \quad $M_{\text{final}}$: Optimal model \Comment{The final model with the best evaluation score}
\State \quad $I_j$: Instruction for the $j$-th task \Comment{$j$-th Task description in natural language}
\State \quad $S_j$: Script generated for $I_j$ \Comment{generated script based on $I_j$}
\State \quad $R_j$: Rendered images for $S_j$ \Comment{Images by rendering $S_j$}
\State \quad $P_j$: Data pair $(I_j, R_j)$ \Comment{Combination of instruction and rendered images}
\State \quad $CF$: Cascade filter for data pair evaluation \Comment{Filters data pairs to ensure quality}
\State \quad $T_i$: Training dataset at iteration $i$ \Comment{Dataset used to train $M_i$}
\State \quad $Loss_i$: Evaluation score for model $M_i$ on validation Set \Comment{Performance on validation Set}

\State \textbf{Initialization:}
\State $i \gets 1$, $M_0 \gets \text{BaseBlenderLLM}$, $S_0 \gets 0$

\While{true} \Comment{Main iterative process}
    \State $T_i \gets \emptyset$ \Comment{Initialize training data for iteration $i$}
    \While{true}
        \State $S_j \sim M_{i-1}(I_j)$ \Comment{Generate script $S_j$ from $M_{i-1}$ using $I_j$}
        \State $R_j = \text{Render}(S_j)$ \Comment{Render images $R_j$ using script $S_j$}
        \State $P_j = (I_j, R_j)$
        \State $CF(P_j) = 
        \begin{cases} 
            \text{Match}, & \text{if $P_{j}$ satisfies filter criteria} \\
            \text{No Match}, & \text{otherwise}
        \end{cases}$ \Comment{Evaluate the data pair using cascade filter}
        \If{$CF(P_j) = \text{Match}$}
            \State $T_i \gets T_i \cup \{P_j\}$ \Comment{Add valid pair to training dataset}
        \Else
            \State Discard $P_j$ \Comment{Ignore invalid data pairs}
        \EndIf
        \If{$|T_i| \geq 2000$}
            \State \textbf{Break} \Comment{Stop collecting data if threshold is met}
        \EndIf
    \EndWhile

    \State $M_i = \text{Train}(M_{i-1}, T_i)$ \Comment{Train model $M_i$ using $M_{i-1}$ and $T_i$}
    \State $Loss_i = \text{Evaluate}(M_i, \text{Validation Set})$ \Comment{Evaluate $M_i$ on Validation Set}
    \If{$Loss_i > Loss_{i-1}$}
        \State $M_{\text{final}} \gets M_{i-1}$ \Comment{Save previous model if score degrades}
        \State \textbf{Break}
    \Else
        \State $M_{i-1} \gets M_i$ \Comment{Update base model for next iteration}
    \EndIf
    \State $i \gets i + 1$ \Comment{Increment iteration counter}
\EndWhile

\State \textbf{Output:} $M_{\text{final}}$
\end{algorithmic}
\end{algorithm*}

\section{Self-improvement Process}
\label{sec:self_improvement}
\subsection{Self-improvement Algorithm}
The algorithm for the Self-improvement process is referenced in Algorithm~\ref{algorithm:Self-improvement_process}.

\subsection{Cascade Filter}
\label{app:cascade_filter}
The classification accuracy of cascade filter is shown in Table~\ref{tab:precision}. Result shows that cascade filter outperforms both single filter.

\begin{table}[]
\centering
\fontsize{8}{10}\selectfont 
\begin{tabular}{cccc}
\hline
\textbf{\begin{tabular}[c]{@{}c@{}}Filters\end{tabular}} &
\textbf{\begin{tabular}[c]{@{}c@{}}Cascade Filter\end{tabular}} &
\textbf{\begin{tabular}[c]{@{}c@{}}Coarse Filter\end{tabular}} &
\textbf{Fine Filter} \\ \hline
\textbf{Precision} &
\textbf{81.8\%} &
61.9\% &
73.3\% \\ \hline
\end{tabular}
\caption{Precision of different Filters. Data deemed acceptable by the Coarse Filter is subsequently processed by the Fine Filter for further verification. This cascaded approach achieves both cost savings and high accuracy.}
\label{tab:precision}
\end{table}

\section{Benchmark}
\label{sec:benchmark}

\subsection{Dimensions for Criteria}

\subsubsection{Object Attributes (\textit{Attr.})}
\textbf{Definition:} This section focuses on evaluating the visual and physical properties of objects, such as shape, color, size, proportion and material characteristics.

\begin{itemize}
    \item \textit{Shape}: \textbf{Shape Accuracy}\\
    Ensure that the objects' shapes align with the instructions, including basic geometries like cubes, spheres, and cylinders.
    \item \textit{Color}: \textbf{Color Representation}\\
    Confirm that the objects' colors precisely match the instructions, including shades, gradients, and lighting effects.
    \item \textit{Size}: \textbf{Size Accuracy}\\
    Check that objects' absolute sizes, such as height, width, and depth, are consistent with the instructions.
    \item \textit{Proportion}: \textbf{Proportion Accuracy}\\
    Ensure the size relationships between different parts of the objects are correct relative to each other.
    \item \textit{Texture}: \textbf{Texture and Surface Detail}\\
    Verify that surface materials like metal, wood, or glass are accurately represented through texture, gloss, or transparency.
\end{itemize}

\subsubsection{Spatial Understanding and Structure (\textit{Spat.})}
\textbf{Definition:} This section evaluates how well the model comprehends and represents the position, relationships, and structure of objects within 3D space.

\begin{itemize}
    \item \textit{Space}: \textbf{Spatial Awareness}\\
    Assess whether the objects' positions and relative relationships within the 3D coordinate system are accurate and logical.
    \item \textit{Contact}: \textbf{Object Contact and Distance}\\
    Verify if the relative distances between objects are reasonable, and whether physical interactions like contact, stacking, or collision are handled correctly.
\end{itemize}

\subsubsection{User Instruction Understanding and Execution (\textit{Inst.})}
\textbf{Definition:} This dimension evaluates how accurately the model interprets and executes the user’s instructions.

\begin{itemize}
    \item \textit{Execute}: \textbf{Execution Accuracy}\\
    Ensure that the objects fully conform to user instructions, including shape, color, size, and material, with no deviations.
\end{itemize}

\subsection{Example for Criteria}
\label{app:criteria_example}
\textbf{Instruction:} \textit{The chair features four cylindrical legs in a deep mahogany color. The seat is circular in a forest green color. Both the backrest and armrests are in the same deep mahogany hue. The height of the legs is 35cm. The height of the armrests is 10cm.} \\
For this instruction, the \textbf{Evaluation Criteria} is:
\begin{itemize}
    \item \textbf{Object Attributes:}
    \begin{itemize}
        \item \textbf{Shape accuracy:}
        \begin{itemize}
            \item \textit{The object in the images is a chair.}
            \item \textit{The chair has four cylindrical legs.}
            \item \textit{The seat is circular.}
            \item \textit{The backrest is rectangular.}
            \item \textit{The armrests are also cylindrical.}
        \end{itemize}
        \item \textbf{Color representation:}
        \begin{itemize}
            \item \textit{The color of the legs is deep mahogany.}
            \item \textit{The seat color is forest green.}
            \item \textit{The backrest color is deep mahogany.}
            \item \textit{The color of the armrests is deep mahogany.}
        \end{itemize}
        \item \textbf{Size:}
        \begin{itemize}
            \item \textit{The height of the legs is 35 cm.}
            \item \textit{The height of the armrests is 10 cm.}
        \end{itemize}
        \item \textbf{Proportion:}
        \begin{itemize}
            \item \textit{The seat is proportionate to the legs.}
            \item \textit{The backrest is at a reasonable height relative to the seat.}
        \end{itemize}
        \item \textbf{Texture and surface detail:}
        \begin{itemize}
            \item \textit{The legs have a smooth wooden texture.}
            \item \textit{The seat may have a fabric texture suitable for upholstery.}
        \end{itemize}
    \end{itemize}
    
    \item \textbf{Spatial Understanding and Structure:}
    \begin{itemize}
        \item \textbf{Three-dimensional spatial awareness:}
        \begin{itemize}
            \item \textit{The legs are positioned correctly for stability.}
            \item \textit{The seat is properly supported by the legs.}
            \item \textit{The backrest is properly supported by the seat.}
            \item \textit{The two armrests are symmetrical.}
        \end{itemize}
        \item \textbf{Object distance and contact:}
        \begin{itemize}
            \item \textit{The legs do not overlap with the seat.}
            \item \textit{There is no gap between the seat and the legs.}
            \item \textit{The backrest connects with the seat at the edge.}
            \item \textit{The armrests are fixed to the backrest and seat.}
        \end{itemize}
    \end{itemize}
    
    \item \textbf{User Instruction Understanding and Execution:}
    \begin{itemize}
        \item \textbf{Instruction execution accuracy:}
        \begin{itemize}
            \item \textit{All specified attributes are accurately represented.}
            \item \textit{There are no deviations from the instructions.}
        \end{itemize}
    \end{itemize}
\end{itemize}

\subsection{Average Number of Criteria across Dimensions}
\label{sec:avg_criteria_num}
\begin{figure}[htbp]
    \centering
    \includegraphics[width=\columnwidth]{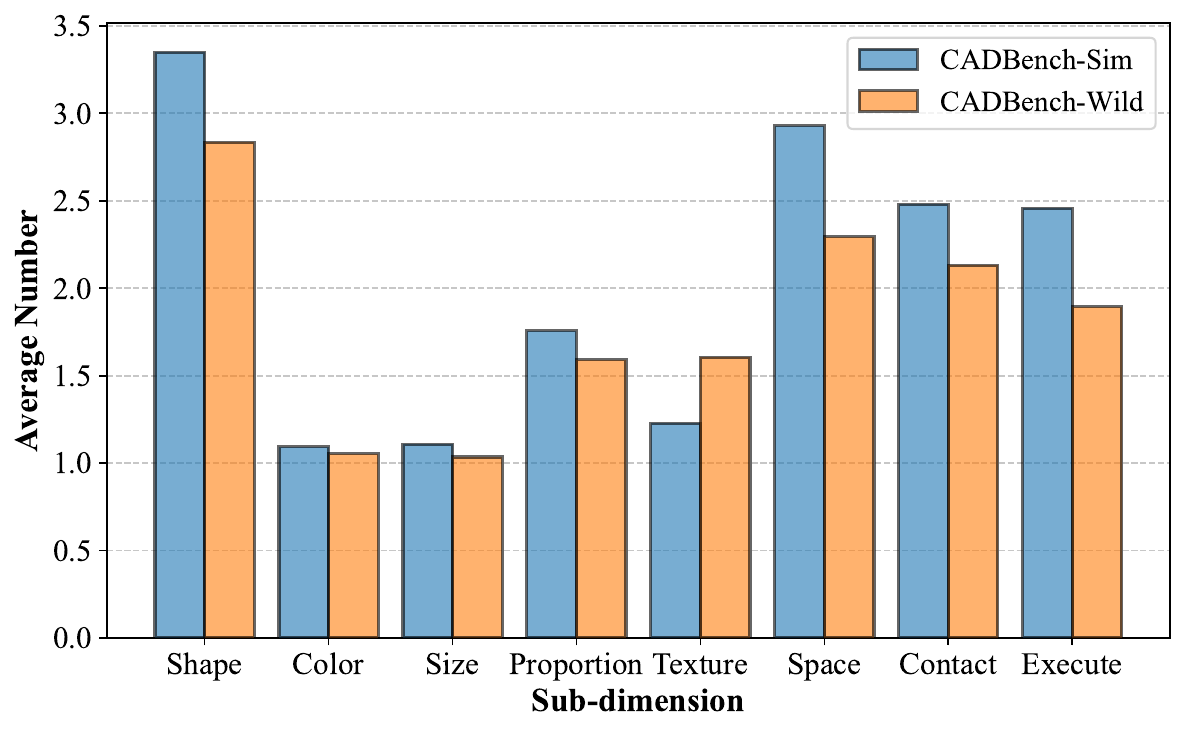}
    \caption{Average number of criteria for each sub-dimension.}
    \label{fig:avg_criteria_num}
\end{figure}

The average number of criteria of each sample across dimensions is shown in Figure~\ref{fig:avg_criteria_num}.

\subsection{Evaluation Metrics}
\label{sec:metrics}

\paragraph{Sub-dimension Scores}
The average score for sub-dimension \(j\) within dimension \(k\), denoted as \(SubDimScore_{k,j}\), is calculated as follows. Here, \(N_{kj}\) represents the total number of criteria in sub-dimension \(j\), and \(S_{kji}\) is the score for the \(i\)-th criterion:

\begin{equation} \small
\label{eq:sub_dimension_score}
SubDimScore_{k,j} = \frac{1}{N_{kj}} \sum_{i=1}^{N_{kj}} S_{kji}
\end{equation}

\paragraph{Dimension Scores}
The average score for a specific dimension \(k\), denoted as \(DimScore_k\), is calculated using Equation \(\ref{eq:dimension_score}\). In this equation, \(N_k\) represents the number of sub-dimensions within dimension \(k\): \

\begin{equation} \small
\label{eq:dimension_score}
DimScore_k = \frac{1}{N_k} \sum_{j=1}^{N_k} SubDimScore_{k,j}
\end{equation}

\paragraph{Overall Scores}
The overall score for a model, denoted as \(Avg.\), is calculated using Equation \(\ref{eq:overall_score}\). In this equation, \(k\) represents the number of dimensions:

\begin{equation} \small
\label{eq:overall_score}
Avg. = \frac{1}{k} \sum_{l=1}^{k} DimScore_{l}
\end{equation}

\paragraph{Syntax Error Rate}

In addition to evaluating the generation quality, we also calculated the syntax error rate ($E_{syntax}$) of the scripts generated by the model. The definition of a syntax error is whether the script generated by the model can successfully produce an image. The $E_{syntax}$ is calculated using Equation \(\ref{eq:error_rate}\). In this equation, $N_{error}$ stands for the number of samples with syntax error, $N_{total}$ stands for the total number of samples:

\begin{equation}
\label{eq:error_rate}
E_{syntax} = \frac{N_{error}}{N_{total}} \times 100\%
\end{equation}

\paragraph{Standard Deviation}

To assess the consistency of the model's outputs, we calculate the $Standard\;Deviation\;(SD)$ of the scores across \(k\) dimensions, as shown in Equation \(\ref{eq:variance}\).

\begin{equation}  \small
  \label{eq:variance}
  SD = \sqrt{\frac{\sum_{l=1}^{k} \left( DimScore_{l} - Avg. \right)^2}{k}}
\end{equation}

\section{Benchmark Evaluation Process}
\label{sec:eval_process}

\begin{figure*}[htbp]
    \centering
    \includegraphics[width=\textwidth]{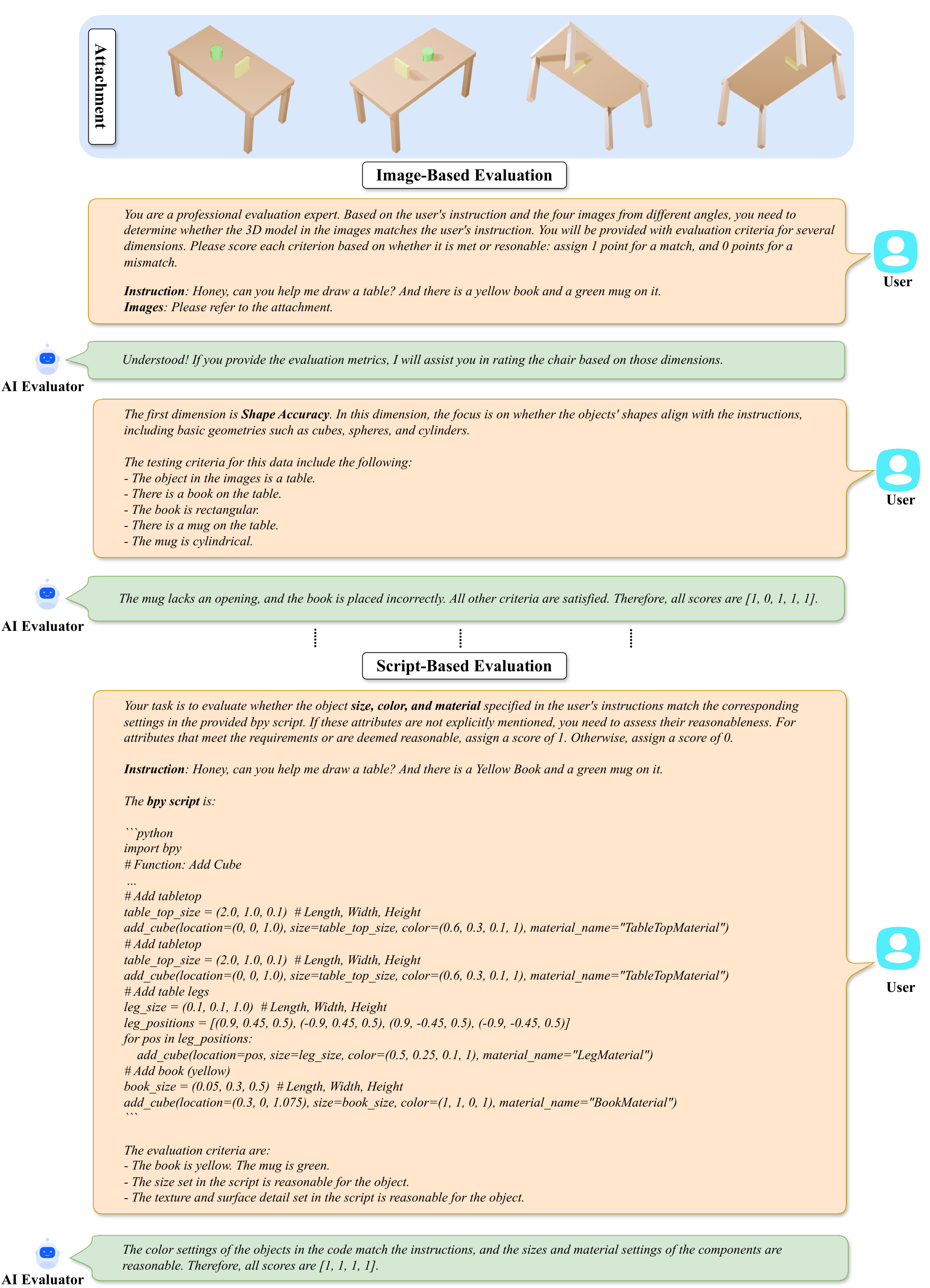}
    \caption{Model Evaluation Process.}
    \label{fig:Benchmark Evaluation Process}
\end{figure*}

For a detailed description of the scoring process, please refer to Figure~\ref{fig:Benchmark Evaluation Process}.

\section{Details of The  Data Generation Pipeline  }
\label{sec:data_gen}

The detailed iterative generation are shown in Algorithm \ref{Algorithm:Iterative Generation and Normalization Process}. The generation prompt is showed in Figure \ref{fig:Prompt Script Generator}.

\section{Human Annotation}

\subsection{Annotation of BlendNet-Human}
\label{sec:BlendNet-Human}

\subsubsection{Objective}
Evaluate the quality of \texttt{<Instruction, Script, Images>} data by ensuring alignment between images, instructions, and scripts to construct the \texttt{BlendNet-Human}.

\subsubsection{Annotation Guidelines}
\begin{itemize}
    \item \textbf{Image-Instruction Alignment:} Images must correspond to the instructions regarding component position, proportion, and specified conditions (e.g., symmetry, rotation, spatial relationships).
    \item \textbf{Script-Instruction Alignment:} Scripts should accurately implement attributes described in the instructions, such as colors, sizes, materials, and other properties not visible in the images.
\end{itemize}

\subsubsection{Annotation Workflow}
\begin{enumerate}
    \item \textbf{Initial Review:} Two annotators independently evaluate each entry, recording \texttt{pass/fail} decisions along with reasons for any failures.
    \item \textbf{Discrepancy Resolution:} A third annotator resolves any disagreements between the initial two annotators.
    \item \textbf{Quality Control:} A QC team reviews 30\% of the data to ensure adherence to guidelines, refining the process based on feedback.
\end{enumerate}

\subsubsection{Team and Results}
\begin{itemize}
    \item \textbf{Annotators:} 12 annotators for initial reviews and 3 annotators for arbitration and quality control.
    \item \textbf{Scale:} Over $10k$ entries were reviewed, resulting in $2k$ entries for \texttt{BlendNet-Human}.
\end{itemize}

\subsection{Annotation of Criteria}
\label{sec:annotationOfCriteria}

\subsubsection{Objective}
Construct the reliable Criteria for \texttt{CADBech} by filtering and modifying $2.5k$ \texttt{<Instruction, Criteria>} pairs to ensure consistency and feasibility.

\subsubsection{Annotation Guidelines}

\paragraph{Instruction Filtering}
\begin{itemize}
    \item \textbf{Relevance and Feasibility:} Instructions must describe feasible and logically sound tasks, excluding ambiguous or unrealistic ones.
    \item \textbf{Material, Surface, and Complexity Constraints:} Instructions with multiple constraints for material, surface details and internal complexity should be simplified to retain only one reasonable requirement.
    \item \textbf{Scope Alignment:} Remove instructions unrelated to the test dataset’s goals.
\end{itemize}

\paragraph{Criteria Validation}
\begin{itemize}
    \item \textbf{Comprehensiveness:} Criteria must cover all dimensions and sub-dimensions.
    \item \textbf{Specificity:} Replace ambiguous terms with measurable criteria.
    \item \textbf{Default for Unspecified Dimensions:} Add default criteria for missing properties (e.g., "color palette should be harmonious").
\end{itemize}

\subsubsection{Annotation Workflow}
\begin{enumerate}
    \item \textbf{Initial Review:} Two annotators independently assess each \texttt{<Instruction, Criteria>} pair, recording decisions and flagging unreasonable data.
    \item \textbf{Discrepancy Resolution:} A third annotator resolves disagreements and finalizes the annotations.
    \item \textbf{Quality Control:} A QC team reviews 30\% of the data to ensure adherence to guidelines, refining the process based on feedback.
\end{enumerate}

\subsubsection{Team and Results}
\begin{itemize}
    \item \textbf{Annotators:} 3 annotators for the annotation process and 1 members in the quality control team.
    \item \textbf{Results:} From the initial $2.5k$ entries, 500 high-quality \texttt{<Instruction, Criteria>} pairs were curated.
\end{itemize}

\subsection{Annotation of Evaluation}
\label{sec:annotationOfScoring}

\subsubsection{Objective}
Obtain human preferences for evaluating the quality of the model's outputs by scoring the results of 200 model responses

\subsubsection{Scoring Guidelines}

\paragraph{Scoring Process}

\begin{itemize}
    \item \textbf{1 point (pass)} if the criterion is satisfied.
    \item \textbf{0 points (fail)} if the criterion is not satisfied.
\end{itemize}

\paragraph{Scoring Criteria}

\subparagraph{1. Image-Based Evaluation}
By comparing the images with the requirements in the instruction, evaluate whether the criteria for all sub-dimensions, except for Color, Size, Texture, and Surface Detail, are met.

\subparagraph{2. Script-Based Evaluation}
By comparing the script with the requirements in the instruction, evaluate whether the criteria for Color, Size, Texture and Surface Detail, are met.

\subparagraph{3. Default Scoring for Unspecified Properties}
\begin{itemize}
    \item Assign \textbf{1 point} if the script logically and harmoniously defines the property.
    \item Assign \textbf{0 points} if the property appears inconsistent or unreasonable.
\end{itemize}

\subsubsection{Annotation Workflow}
\begin{enumerate}
    \item \textbf{Data Assignment:} Annotators are assigned all of \texttt{<Instruction, Script, Images>} entries (four images per entry).
    \item \textbf{Scoring and Justification:} Annotators score each criterion and provide explanations for any failing scores.
    \item \textbf{Quality Control:} A QC team reviews 30\% of the data to ensure compliance with guidelines, refining the process based on feedback.
\end{enumerate}

\subsubsection{Team and Results}
\begin{itemize}
    \item \textbf{Annotators:} 3 scoring annotators and 1 quality control annotators.
    \item \textbf{Results:} The \textit{kappa} value, calculated to reflect the consistency between human evaluators, is 0.883.
\end{itemize}

\section{Baseline LLMs}
\label{sec:baselines}
Details about the baseline LLMs are shown below:

\begin{itemize}
    \item \textbf{o1-Preview} \citep{o1preview2024}: O1-Preview is a version of OpenAI's O1 model. It provides enhanced efficiency and accuracy for diverse applications, delivering high-performance results with optimized capabilities.
    \item \textbf{GPT-4 turbo} \citep{openai2023gpt4}: GPT-4 Turbo is a version of OpenAI's GPT-4 model. It offers improved performance in responses for a wide range of applications.
    \item \textbf{Claude3.5-sonnet} \citep{anthropic2023claude}: A model developed by Anthropic, known for its safety and alignment features in language generation tasks.
    \item \textbf{GPT-4o} \citep{openai2024gpt4o}:GPT-4o is a language model developed by OpenAI that can generate human-like text based on the input it receives.
    \item \textbf{BlenderGPT} \citep{blendergpt2023}: A model developed by Aarya and Flip Phillips, which allows user to use natural language commands to control Blender. It leverages GPT-3.5 \citep{OpenaiGPT3} or GPT-4 \citep{openai2023gpt4} to generate corresponding {\tt bpy} scripts based on user-defined prompts for rendering 3D models.
    \item \textbf{Gemini-1.5-pro} \citep{googlegemini2024}: Gemini 1.5 is an advanced AI language model developed by Google DeepMind.
    \item \textbf{DeepSeek-V2.5} \citep{deepseek2024}: DeepSeek-V2.5 is an advanced language model designed for information retrieval tasks, optimized for search accuracy and efficiency across large datasets.
    \item \textbf{Qwen-2.5-Coder-7B-Instruct} \citep{qwen25_2024}: Qwen2.5-Coder is the latest series of Code-Specific Qwen large language models
    \item \textbf{Qwen-2.5} \citep{qwen25_2024}: Qwen-2.5 is a versatile language model that excels in natural language understanding and generation, providing improved context comprehension and response accuracy.
    \item \textbf{LLaMA3} \citep{touvron2023llama}: The latest version of the LLaMA model, which has been fine-tuned for a variety of natural language processing tasks.
    \item \textbf{Mistral-7B-Instruct-V0.3} \citep{mistral_2024}: Mistral-7B-Instruct-V0.3 is a highly scalable model known for its performance in both text generation and comprehension tasks, utilizing 8-layer attention mechanisms with a 7B parameter architecture for enhanced processing.
    \item \textbf{CodeLLaMa-7B-Instruct} \citep{codellama2024}: Code Llama is a collection of pretrained and fine-tuned generative text models ranging in scale from 7 billion to 34 billion parameters.

\end{itemize}

\section{Performance on Sub-Dimensions}
\label{sec:Performance of LLMs on Sub-Dimension}

\begin{figure*}[t]
    \centering
    \includegraphics[width=\textwidth]{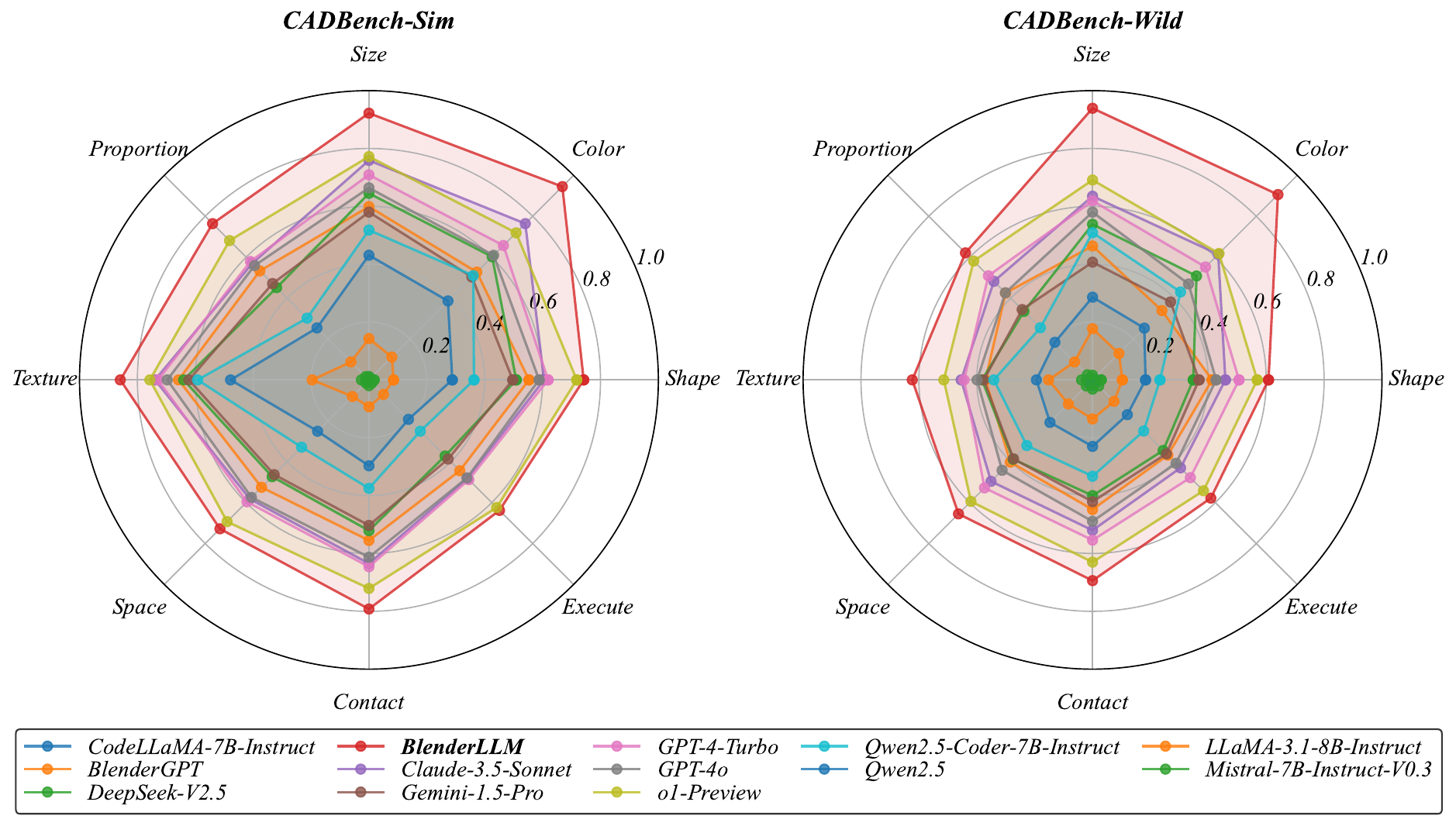}
    \caption{Performance of different LLMs on Sub-Dimensions}
    \label{fig:performance_sub_dimension}
\end{figure*}

The performance of different LLMs on Sub-Dimension is shown in Figure~\ref{fig:performance_sub_dimension}.

\section{Visual Performance of Different Models}
\label{sec:The Visual Examples of the Performance of Different Models}

The Visual Examples of the Performance of Different Models are shown in Table~\ref{tab:sample_performance}.

\section{Visual Performance of Different Training strategy}
\label{sec:The Visual Examples of the Performance of Different Training strategy}

The Visual Examples of the Performance of Different Training strategy are shown in Table~\ref{tab:sample_strategy}.

\section{Characteristics of Annotators}

The annotators involved in this study possess the following characteristics:

\begin{itemize}
    \item Bachelor's degree in one of the following fields: Computer Science, Data Science, Business Administration, English, Music, or Biological Sciences.
    \item Full English instruction during their academic education.
\end{itemize}

\section{AI Assistant}
Some of the text has been polished and revised by \texttt{GPT-4}, but the main part is completed by humans.

\onecolumn

\begin{longtblr}[
  caption = {The Visual Examples of the Performance of Different Models}, 
  label = {tab:sample_performance}, 
]{
  width = \textwidth,
  colspec = {Q[80]Q[100]Q[38]Q[100]Q[38]Q[100]Q[38]}, 
  cells = {c}, 
  rows = {m}, 
  cell{1}{1} = {r=5}{m}, 
  cell{1}{2} = {c=6}{0.24\linewidth}, 
  cell{2}{2} = {c=2}{0.24\linewidth}, 
  cell{2}{4} = {c=2}{0.24\linewidth}, 
  cell{2}{6} = {c=2}{0.24\linewidth}, 
  cell{3}{2} = {c=6}{0.24\linewidth}, 
  cell{4}{2} = {c=2}{0.24\linewidth}, 
  cell{4}{4} = {c=2}{0.24\linewidth}, 
  cell{4}{6} = {c=2}{0.24\linewidth}, 
  vlines, 
  hline{1,6-19} = {-}{}, 
  hline{2-5} = {2-7}{}, 
}
\textbf{Models } & \textbf{Dimension} &  &  &  &  & \\
 & \textit{Atrr.} &  & \textit{Spat.} &  & \textit{Inst.} & \\
 & \textbf{Instruction} &  &  &  &  & \\
 & \small \textit{Create a 3D model of a \uline{burger}. It consists of a sesame seed bun, a beef patty, a slice of cheese, lettuce, tomato, and pickles.} &  & \small \textit{I need better \uline{lighting }on my desk and want a functional and stylish desk lamp, would you be able to give me some functional and stylish construction?} & 
 & \small \textit{Design a 3D model of a \uline{Celtic knot.} The knot should be intricate, with interlocking loops and a continuous pattern. Ensure the design is symmetrical and has a traditional Celtic feel.} & \\
 & \textbf{Images} & \textbf{Scores}  & \textbf{Images} & \textbf{Scores} & \textbf{Images} & \textbf{Scores} \\
BlenderLLM & \raisebox{-0.5\height}{\includegraphics[scale=0.35]{burger/blenderllm.pdf}} & \textbf{1.0} & \raisebox{-0.5\height}{\includegraphics[scale=0.3]{lamp/lamp_blenderllm.pdf}} & \textbf{1.0} & \raisebox{-0.5\height}{\includegraphics[scale=0.25]{knot/knot_blenderllm.pdf}} & \textbf{1.0} \\
o1-Preview & \raisebox{-0.5\height}{\includegraphics[scale=0.35]{burger/o1.pdf}} & 0.8 & \raisebox{-0.5\height}{\includegraphics[scale=0.3]{lamp/lamp_o1.pdf}} & 0.4 & \raisebox{-0.5\height}{\includegraphics[scale=0.3]{knot/knot_o1.pdf}} & 0 \\
GPT-4-Turbo & \raisebox{-0.5\height}{\includegraphics[scale=0.3]{burger/turbo.pdf}} & 0.8 & \raisebox{-0.5\height}{\includegraphics[scale=0.3]{lamp/lamp_turbo.pdf}} & 0.4 & \textit{Syntax Error} & 0 \\
Claude-3.5-Sonnet & \textit{Syntax Error} & 0 & \raisebox{-0.5\height}{\includegraphics[scale=0.3]{lamp/lamp_claude.pdf}} & 0.2 & \textit{Syntax Error} & 0 \\
GPT-4o & \raisebox{-0.5\height}{\includegraphics[scale=0.3]{burger/4o.pdf}} & 0.6 & \raisebox{-0.5\height}{\includegraphics[scale=0.3]{lamp/lamp_gpt-4o.pdf}} & 0.8 & \raisebox{-0.5\height}{\includegraphics[scale=0.3]{knot/knot_gpt-4o.pdf}} & 0.5\\
BlenderGPT & \raisebox{-0.5\height}{\includegraphics[scale=0.3]{burger/blendergpt.pdf}} & 0.5 & \raisebox{-0.5\height}{\includegraphics[scale=0.3]{lamp/lamp_blendergpt.pdf}} & 0.8 & \textit{Syntax Error} & 0\\
Gemini-1.5-Pro & \raisebox{-0.5\height}{\includegraphics[scale=0.3]{burger/gemini.pdf}} & 0.5 & \raisebox{-0.5\height}{\includegraphics[scale=0.3]{lamp/lamp_gemini.pdf}} & 0.2 & \textit{Syntax Error} & 0\\
DeepSeek-V2.5 & \textit{Syntax Error} & 0 & \raisebox{-0.5\height}{\includegraphics[scale=0.35]{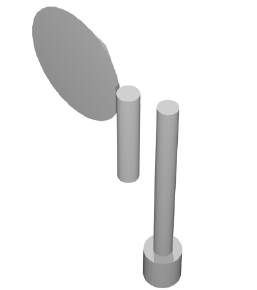}} & 0.2 & \textit{Syntax Error} & 0\\
Qwen2.5-Coder-7B-Instruct & \raisebox{-0.5\height}{\includegraphics[scale=0.3]{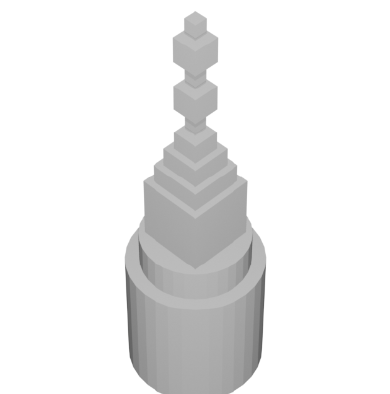}} & 0.2 & \raisebox{-0.5\height}{\includegraphics[scale=0.3]{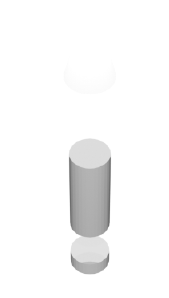}} & 0 & \textit{Syntax Error} & 0\\
Qwen2.5 & \raisebox{-0.5\height}{\includegraphics[scale=0.3]{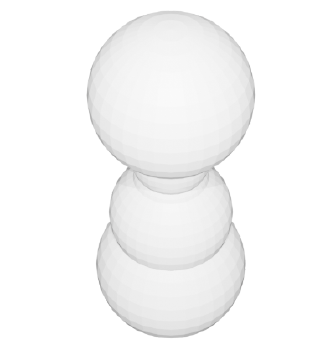}} & 0.2 & \textit{Syntax Error} & 0 & \textit{Syntax Error} & 0\\
LLaMA-3.1-8B-Instruct & \textit{Syntax Error} & 0 & \textit{Syntax Error} & 0 & \textit{Syntax Error} & 0\\
Mistral-7B-Instruct-V0.3 & \textit{Syntax Error} & 0 & \textit{Syntax Error} & 0 & \textit{Syntax Error} & 0\\
CodeLLaMA-7B-Instruct & \textit{Syntax Error} & 0 & \textit{Syntax Error} & 0 & \textit{Syntax Error} & 0
\end{longtblr}

\twocolumn

\onecolumn

\begin{longtblr}[
  caption = {The Visual Examples of the Performance of Different Training strategy},
  label = {tab:sample_strategy},
]{
  width = \textwidth,
  colspec = {Q[313]Q[313]Q[313]},
  row{even} = {c},
  row{3} = {c},
  cell{1}{1} = {c=3}{0.943\linewidth},
  hlines,
  vlines,
}
\textbf{Instruction:} \textit{Can you help me to draw a chair? It has regular legs, a square seat and a square back with yellow stripes.} &  & \\
Self-improvement Training & Epoch Accumulation Training & Predefined Incremental Training\\
\includegraphics[scale=0.3]{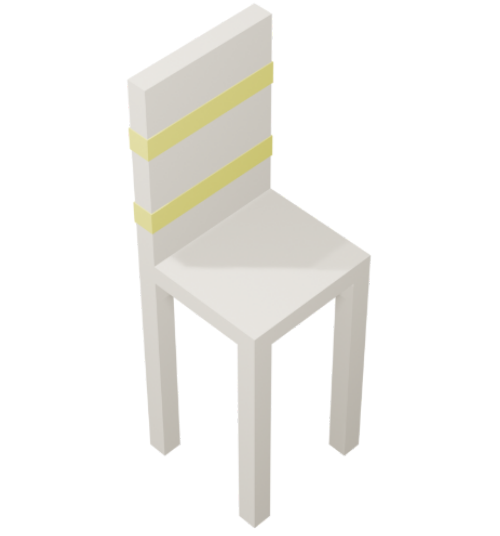} & \includegraphics[scale=0.3]{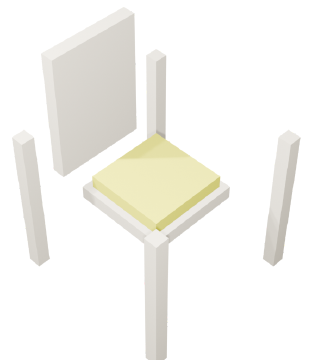} & \includegraphics[scale=0.3]{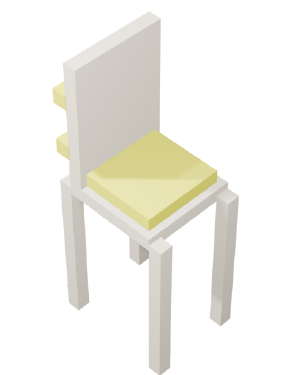}\\
\includegraphics[scale=0.3]{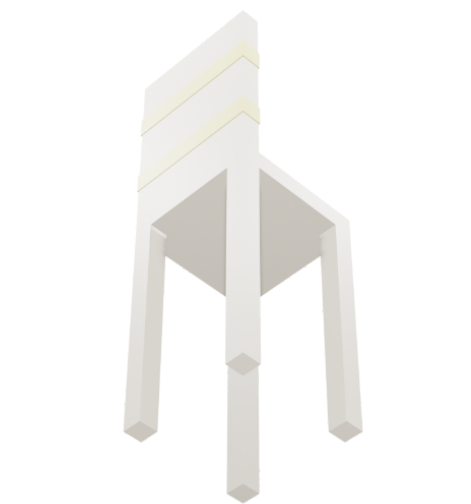} & \includegraphics[scale=0.3]{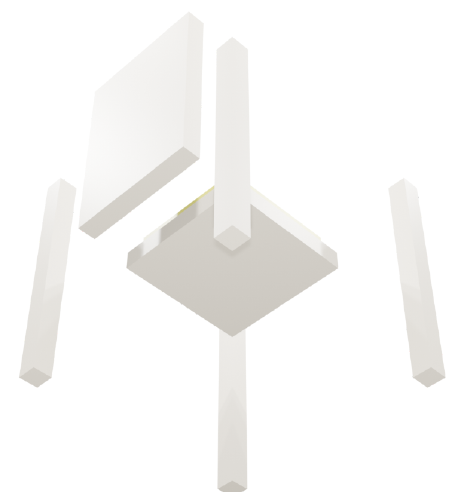} & \includegraphics[scale=0.3]{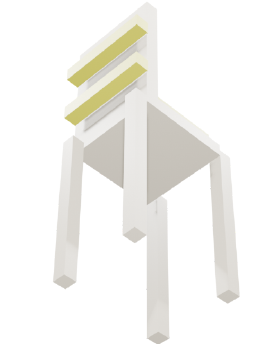}
\end{longtblr}
\twocolumn

\end{document}